\documentclass[final,3p,times,authoryear]{elsarticle}

\usepackage{amssymb}

\journal{Ocean Modelling}

\usepackage{color}
\usepackage{amsmath}

\newcommand{\dy}{\partial}
\newcommand\ddy[2]{\frac{\dy#1}{\dy#2}}

\newcommand{\ex}{\mathrm{e}}

\newcommand{\grad}{\nabla}

\newcommand{\Fb}{\boldsymbol{F}}
\newcommand{\Gb}{\boldsymbol{G}}
\newcommand{\Hb}{\boldsymbol{H}}

\newcommand{\Rb}{\boldsymbol{R}}

\newcommand{\nb}{\boldsymbol{n}}

\newcommand{\ub}{\boldsymbol{u}}

\newcommand{\xb}{\boldsymbol{x}}

\newcommand{\zb}{\boldsymbol{z}}

\definecolor{dark-green}{rgb}{0,0.5,0} 

\usepackage{rotating}
\usepackage{relsize}

\newcommand{\kappaGM}{\kappa_{\scriptsize\mbox{gm}}}

\usepackage{listings}
\usepackage{textcomp}
\definecolor{grey}{rgb}{0.5,0.5,0.5}
\definecolor{mauve}{rgb}{0.58,0,0.82}
\definecolor{darkgreen}{rgb}{0,0.3,0}
\definecolor{darkred}{rgb}{0.8,0,0}
\begin{document}

\begin{frontmatter}



\title{A new gauge-invariant method for diagnosing eddy diffusivities}


\author[edinburgh]{J. Mak}
\author[edinburgh]{J. R. Maddison}

\address[edinburgh]{School of Mathematics and Maxwell Institute for Mathematical
Sciences, University of Edinburgh, Edinburgh, EH9 3FD, United Kingdom}

\author[oxford]{D. P. Marshall}

\address[oxford]{Department of Physics, University of Oxford, Oxford, OX1 3PU,
United Kingdom}

\begin{abstract}
Coarse resolution numerical ocean models must typically include a
parameterisation for mesoscale turbulence. A common recipe for such
parameterisations is to invoke down-gradient mixing, or diffusion, of some
tracer quantity, such as potential vorticity or buoyancy. However, it is well
known that eddy fluxes include large rotational components which necessarily do
not lead to any mixing; eddy diffusivities diagnosed from unfiltered fluxes are
thus contaminated by the presence of these rotational components. Here a new
methodology is applied whereby eddy diffusivities are diagnosed directly from
the eddy force function. The eddy force function depends only upon flux
divergences, is independent of any rotational flux components, and is inherently
non-local and smooth. A one-shot inversion procedure is applied, minimising the
mis-match between parameterised force functions and force functions derived from
eddy resolving calculations. This enables diffusivities associated with the eddy
potential vorticity and buoyancy fluxes to be diagnosed. The methodology is
applied to multi-layer quasi-geostrophic ocean gyre simulations. It is found
that: (i) a strictly down-gradient mixing scheme has limited success in reducing
the mis-match compared to one with no sign constraint on the diffusivity; (ii)
negative signals of diffusivities are prevalent around the time-mean jet; (iii)
there is some indication that the magnitude of the diffusivity correlates well
the eddy energy. Implications for parameterisation are discussed in light of
these diagnostic results.
\end{abstract}


\end{frontmatter}


\section{Introduction}

A key challenge in ocean modelling is to improve the representation of turbulent
mesoscale eddies in the models used for long-range climate projections, for
which routine explicit resolution of the turbulent eddy fluxes is unlikely for
the next few decades. Turbulence closures are very commonly based upon mixing
principles: small scale ``eddy'' dynamics should, on average, lead to mixing of
large scale ``mean'' fields. In the atmosphere and ocean this principle is
typically applied to the potential vorticity (PV), via the introduction of an
eddy PV diffusivity \citep{Green70, Marshall81}. For example, it is well-known
that PV tends to be mixed in closed ocean gyres \citep{RhinesYoung82}. More
generally, eddy enstrophy is dissipated on small scales, and correspondingly
eddy PV fluxes lead to a net generation of eddy enstrophy on average, i.e., eddy
PV fluxes are oriented down-gradient in a domain integral sense. The success of
a down-gradient PV parameterisation, therefore, is dependent upon the degree to
which this mixing principle, which must hold in a domain integral sense, is
valid in a local sense.

Locally, however, eddy enstrophy may be significantly transported by mean and
eddy advection, and also be influenced by local forcing. While the eddy PV
fluxes are oriented down-gradient on average, there is in general no constraint
on their local orientation. In particular, the eddy PV fluxes can be separated
into advective, rotational, and residual components
\citep[e.g.][]{MedvedevGreatbatch04}, with only the latter leading to local
mixing. Considerations of the eddy enstrophy budget allows the advective
component to be defined in terms of the mean advection of enstrophy
\citep{MarshallShutts81, McDougallMcIntosh96, Nakamura98}, or the mean and eddy
advection (\citealt{MedvedevGreatbatch04}; see also \citealt{Eden-et-al07} for
further generalisation). It is known, moreover, that eddy PV fluxes can contain
large rotational components \citep[e.g.,][]{Griesel-et-al09}, which have no
direct effect on the mean dynamics and necessarily lead to no mixing. Formally
the dynamics is invariant under the addition of an arbitrary rotational gauge to
the eddy PV flux (which vanishes under the divergence). Rotational PV flux
components can be removed via a Helmholtz decomposition, although such a
decomposition in a bounded domain is non-unique, as there is freedom in the
specification of the boundary conditions \citep{FoxKemper-et-al03}. These issues
complicate the diagnosis of eddy diffusivities. Recently,
\citet{Maddison-et-al15} have shown that, at least for quasi-geostrophic eddy PV
fluxes, one can define an \emph{eddy force function} which simultaneously
defines the forcing of the mean flow by the eddies and a unique divergent
component of eddy PV fluxes. Moreover, in a simply connected domain, the
divergent PV flux thus defined is optimal, in the sense that it has minimum
magnitude (specifically minimal domain integrated squared magnitude, or
equivalently minimal $L^2$ norm).

In this article an alternative gauge-invariant diagnostic approach is proposed
which simultaneously avoids any ambiguity associated with the presence of
rotational fluxes, and also takes into account the inherent non-locality of the
dynamic influence of eddy PV fluxes. This is achieved via an optimisation
procedure. Specifically, given mean fields computed in an eddy resolving
calculation, together with a candidate eddy parameterisation, an associated
parameterised eddy force function can be calculated via the solution of a
Poisson equation. The approach is thus inherently non-local, in the sense that
parametised eddy force function depends upon the parameterisation itself through
an inverse elliptic operator. A parameterisation quality cost function is
defined via a measure of the mis-match between this parameterised eddy force
function, and the eddy force function diagnosed from the original eddy resolving
calculation. Eddy diffusivities can then be diagnosed via the solution of an
inverse problem: seeking the diffusivity which minimises the mis-match between
the parameterised and diagnosed force functions. Ill-posedness of the inversion
is treated via the introduction of an additional regularisation, acting to
smooth the diagnosed diffusivity. The partial differential equation constrained
optimisation problem itself is solved via a one-shot approach \citep[][\S
2.2]{Gunzburger-control}, with the associated optimality system constructed and
solved via the use of the FEniCS automated code generation system \citep[see
e.g.][]{Logg-et-al-FENICS}.

The layout of this article is as follows. In \S2 details regarding the eddy
force function are reviewed. The optimisation problem for eddy PV diffusivities
is formulated, and the numerical implementation is described. In \S3 eddy
diffusivities associated with PV and buoyancy mixing are diagnosed using this
procedure; the diagnostic is applied to model data based upon the ocean gyre
calculations described in \citet{Maddison-et-al15}, computed using a three-layer
quasi-geostrophic finite element model. The diagnostic calculations are repeated
using data from a higher resolution, five-layer finite difference calculation in
\S4. The paper concludes in \S5, and consequences for geostrophic eddy
parameterisation are considered.


\section{Formulation}

Throughout this article we limit consideration to mesoscale dynamics, and
specifically to the quasi-geostrophic (QG) equations. The fundamental principle
applied is to formulate a method for the diagnostic calculation of eddy
diffusivities in a way that is independent of any rotational eddy flux
components -- that is, to formulate a gauge invariant diagnostic. This is
tackled by constructing a constrained optimisation problem, whereby a
parameterised diffusivity is diagnosed by minimising a measure of the mis-match
between parameterised and diagnosed eddy force functions, which depend only on
the PV flux divergence. The critical step is defining an appropriate measure of
the mis-match in order to avoid undue sensitivity to small scale noise in the
divergence field.

The optimisation problem used to achieve this is outlined in
\S\ref{sec:optimisation}. A measure of this mis-match is defined via the use of
the eddy force function, introduced in \cite{Maddison-et-al15}. For
completeness, mathematical background regarding the eddy force function is
provided in \S\ref{sec:force-func}. A parameterised eddy force function is
computed from parameterised eddy fluxes via the solution of an elliptic problem.
This leads naturally to the formulation of a PDE constrained optimisation
problem which diagnoses the diffusivity. Implementation details are provided in
\S\ref{sec:formulation}.


\subsection{Unconstrained optimisation problem for eddy diffusivities}\label{sec:optimisation}

The mean QGPV equation takes the form
\begin{equation}\label{eq:qgpv}
  \ddy{\overline{q}}{t} +
    \grad \cdot \left( \overline{\ub}_g \overline{q}\right) 
  = -\grad \cdot \Fb + \overline{Q},
\end{equation}
where $q$ is the PV, $\ub_g$ is the non-divergent geostrophic velocity, $\Fb =
\overline{\ub_g' q'}$ is the eddy PV flux, $Q$ represents all forcing and
dissipation, $\grad$ is the horizontal gradient operator, and $t$ is time. An
overline denotes the mean, a prime the derivation from the mean, and the mean
operator is a Reynolds operator which commutes with all relevant derivatives
(cf. \citealt{MaddisonMarshall13}).

Consider a down-gradient PV parameterisation. If the mean PV gradient is
non-zero, the eddy PV flux $\Fb$ may expressed as
\begin{equation}
  \Fb = - \kappa \grad \overline{q} - \sigma \hat{\zb} \times \grad \overline{q} + \Rb,
\end{equation}
where $\kappa$ is the PV diffusivity and $\sigma$ a skew-diffusion coefficient
(equal to a stream function associated with eddy-induced advection; see
\citealt{Vallis-GFD}, \S10.6.3), and $\Rb$ is any non-divergent flux. In general
$\Rb$ is the sum of rotational and harmonic flux and, as it vanishes under the
divergence, it has no direct influence on the mean dynamics. Taking the scalar
product with the mean PV gradient leads to a definition for the local PV
diffusivity
\begin{equation}\label{eq:kappa_noise}
  \kappa = -\frac{\left( \Fb - \Rb \right) \cdot \grad \overline{q}}{\left| \grad \overline{q}\right|^2}.
\end{equation}
The central issue is ambiguity in the definition of $\Rb$. For example an
approximately rotational component of $\Fb$ may be associated with local
advection of enstrophy, rather than generation of enstrophy and hence not
correspond to local irreversible mixing \citep{MarshallShutts81}. The mean
dynamics are invariant under any choice of the non-divergent gauge $\Rb$, but
the diffusivity as defined by equation \eqref{eq:kappa_noise} is not. Moreover a
diffusivity field diagnosed in this way may be extremely noisy
\citep[cf.][]{NakamuraChao00}, as shown by a sample calculation employing this
approach using the simulation data presented in \S3. In this diagnostic
calculation there are regions of large negative diffusivity, which suggests the
presence of strong eddy backscatter (conversion of eddy to mean enstrophy).
These negative diffusion regions may be due to pollution of the diagnostic by
significant non-divergent eddy PV fluxes and, in this sense, be entirely
artificial. Critically, this direct approach fails to unambiguously identify the
regions and magnitude of irreversible mixing due to the eddies.

\begin{figure}[tbh]
\begin{center}
  \includegraphics[width=0.8\textwidth]{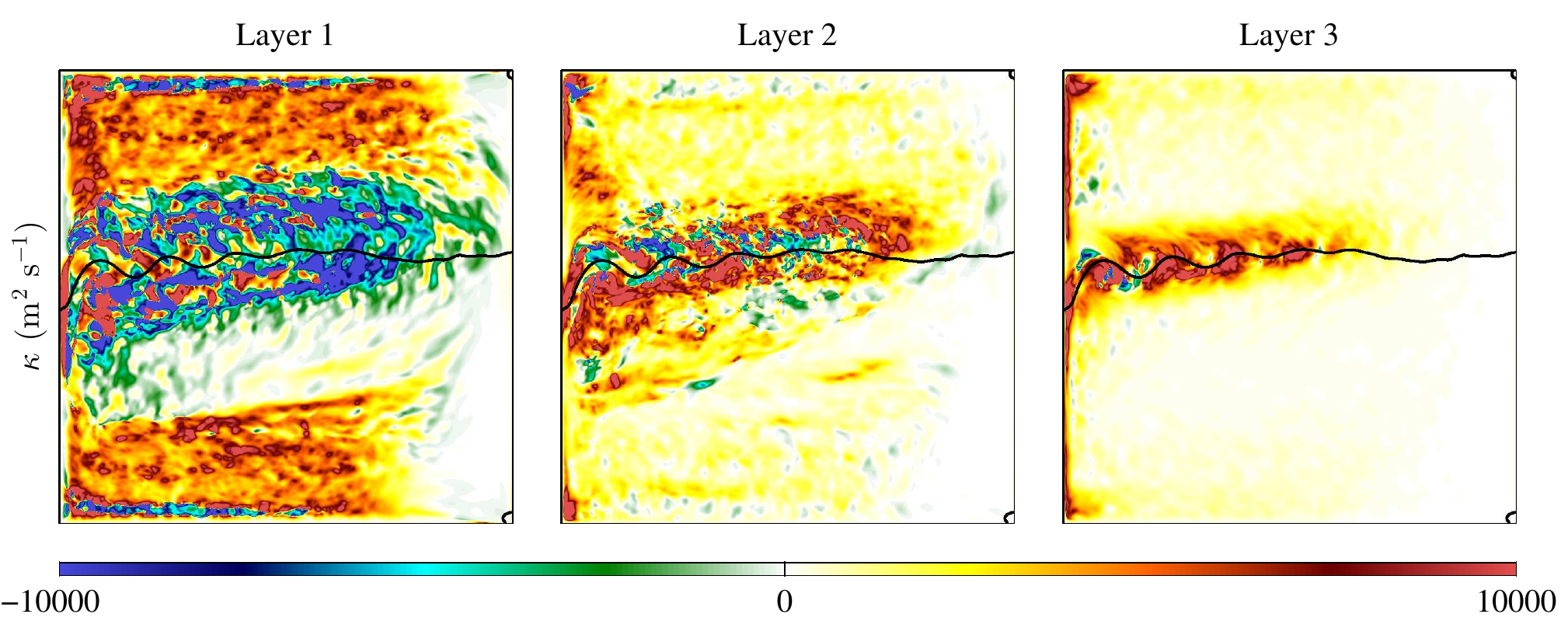}
  \caption{Local eddy PV diffusivities $\kappa$ (in units of $\mathrm{m}^2\
  \mathrm{s}^{-1}$), obtained from \eqref{eq:kappa_noise} with $\Rb = 0$ using
  the simulation data detailed in \S3. Note that the colour scale is saturated.}
  \label{fig:kappaPV_data}
\end{center}
\end{figure}

The gauge freedom may be formally addressed using a horizontal Helmholtz
decomposition
\begin{equation}\label{eq:helmholtz}
  \Fb = -\nabla \tilde{\Psi} + \hat{\zb} \times \nabla \tilde{\Phi} +
  \tilde{\Hb},
\end{equation}
where the first term is the \emph{divergent} component, the second the
\emph{rotational} component, and the third a harmonic component. Since only the
former is directly dynamically active, a down-gradient parameterisation may now
alternatively be expressed in terms of the divergent component
\begin{equation}
  -\nabla \tilde{\Psi} = - \kappa_{\tilde{\Psi}} \nabla \overline{q} -
  \sigma_{\tilde{\Psi}} \hat{\zb} \times \nabla \overline{q},
\end{equation}
leading to an alternative definition for the local PV diffusivity
\begin{equation}\label{eq:kappa_div}
  \kappa_{\tilde{\Psi}} = \frac{\nabla \tilde{\Psi} \cdot \nabla
\overline{q}}{\left| \nabla \overline{q} \right|^2}.
\end{equation}
Given any divergent component of the eddy PV fluxes one can thus define a local
diffusivity (if the mean PV gradient is non-zero) via equation
\eqref{eq:kappa_div}. However, in the presence of boundaries, the divergent
component is non-unique due to freedom in the choice of boundary conditions for
the potential $\tilde{\Psi}$ \citep{FoxKemper-et-al03}. Hence the eddy
diffusivity as defined by equation \eqref{eq:kappa_div} is still not uniquely
defined.

This ambiguity can be resolved by instead defining a PV diffusivity directly
from the eddy PV flux divergence. Specifically, if a pure down-gradient eddy PV
flux is postulated, that is,
\begin{equation}
  \Fb \approx -\kappa \grad \overline{q},
\end{equation}
then an optimal PV diffusivity can be defined by seeking the spatially varying
function $\kappa \left( \xb \right)$ such that the cost function
\begin{equation}\label{eq:mis-match}
  \mathcal{J} \left( \kappa \right) = \left\| \grad \cdot \left( -\kappa
  \grad\overline{q} - \Fb \right) \right\|^2
\end{equation}
is minimised. This defines a best-fit for the diffusivity to the eddy flux
divergence. The specific norm, which has been left unspecified for the moment,
is a key ingredient in the definition of this optimal diffusivity. Note that a
perfect match is not typically to be expected and, moreover, the inversion may
be highly ill-conditioned (or even ill-posed).

The norm appearing in \eqref{eq:mis-match} has an important impact on the
structure of the resulting optimal diffusivity. It is clear for example that a
simple $L^2$ norm, which leads to
\begin{equation}\label{eq:mis-match_force_function}
  \mathcal{J} \left(\kappa\right) 
    = \left\|\grad\cdot\left(-\kappa \overline{q} - \Fb \right)\right\|_{L^2}^2
  = \int_\Omega \left[\grad\cdot\left(-\kappa\overline{q}-\Fb\right)\right]^2
    \, \mathrm{d}\Omega,
\end{equation}
where $\Omega$ is the horizontal domain, will lead to difficulties. The
divergence of a flux is an inherently noisy quantity, and hence an attempt to
optimise the diffusivity to match local structure in the eddy PV flux divergence
is likely to be problematic.


\subsection{Eddy force function}\label{sec:force-func}

A starting point is to first resolve the non-uniqueness in the definition of the
eddy PV flux decomposition in equation \eqref{eq:helmholtz}. This is addressed
in \citet{Maddison-et-al15} by noting a relationship between rotational momentum
tendencies and divergent eddy PV fluxes, which is briefly summarised.

First, it is noted that the QG momentum equation may be written
\begin{equation}
  \ddy{\overline{\ub_g}}{t} = -\hat{\zb}\times\Gb.
\end{equation}
Since the geostrophic velocity is non-divergent, $\grad\cdot\overline{\ub_g}=0$,
so $-\hat{\zb}\times\Gb$ defines a unique rotational momentum tendency. After
taking a horizontal curl, $\Gb$ may be identified as a horizontally divergent PV
flux. Introducing the stream function $\overline{\psi}$ such that
$\overline{\ub_g} = \hat{\zb}\times\grad\overline{\psi}$ yields
$\dy\overline{\ub_g}/\dy t = \hat{\zb} \times\grad\Psi$, from which it follows
that
\begin{equation}
  \Psi = \ddy{\overline{\psi}}{t} + c(z,t),
\end{equation}
Following \cite{MarshallPillar11}, $\Psi$ is a stream function tendency
or \emph{force function}. In a simply connected domain with no-normal-flow
boundary conditions $\overline{\psi}$ is a (horizontal) constant on all
boundaries, and so $\Psi$ inherits a Dirichlet boundary condition. Subject to an
appropriate choice of $c(z,t)$ (noting that any other choice vanishes under the
horizontal divergence), $\Psi$ satisfies a homogeneous Dirichlet boundary
condition. Insisting that the force function decomposition procedure is linear
then implies that a force function associated with any single momentum tendency
inherits a homogeneous Dirichlet boundary condition.

In particular, the eddy force function $\Psi_e$ is related to the eddy PV flux
$\Fb = \overline{\ub'q'}$ by
\begin{equation}\label{eq:force-function}
  \Fb = -\grad \Psi_e + \hat{\zb}\times\grad\Phi_e +
  \boldsymbol{H}_e,
\end{equation}
where $\Psi_e$ is the solution of a Poisson equation
\begin{equation}\label{eq:ff_poisson}
  \nabla^2 \Psi_e = -\grad \cdot \Fb
\end{equation}
subject to homogeneous Dirichlet boundary conditions.

The eddy force function has a number of important properties. First, since only
the divergence of the eddy PV flux appears in the force function equation
\eqref{eq:ff_poisson}, it is independent of rotational eddy fluxes. Moreover the
eddy force function is inherently smooth. It is shown that, in a simply
connected domain, the eddy force function has minimal $H^1_0$ semi-norm, that
is, it is a solution to the Poisson equation \eqref{eq:ff_poisson} for which the
mean square gradient is minimised \cite[][\S2 and Appendix A]{Maddison-et-al15}.
Note that the eddy force function depends non-locally upon the eddy fluxes ---
it is related to the flux divergence through an inverse elliptic operator. This
suggests that the mis-match function \eqref{eq:mis-match} be defined in terms of
the mis-match between the eddy force function implied by a parameterisation, and
the eddy force function diagnosed from data; that is,
\begin{equation}\label{eq:unconstrained_function}
  \mathcal{J} \left( \kappa \right)
  = \left\| \Psi_p \left( \kappa \right) - \Psi_e \right\|^2
\end{equation}
where $\Psi_p$ is the parameterised eddy force function computed via
\begin{equation}\label{eq:ffp_poisson}
  \grad^2 \Psi_p = \grad \cdot \left( \kappa \grad \overline{q} \right)
\end{equation}
subject to homogeneous Dirichlet boundary conditions. 

There is now freedom in the definition of the norm appearing in
\eqref{eq:unconstrained_function}. A simple choice is to define this to be
the $L^2$ norm, leading to
\begin{equation}
  \mathcal{J} \left( \kappa \right)
  = \left\|\grad\cdot\left(-\kappa \grad\overline{q} - \Fb \right)\right\|^2 
  = \left\|\Psi_p \left( \kappa \right) - \Psi_e \right\|_{L^2}^2
  = \int_\Omega \left[ \Psi_p \left(\kappa\right) - \Psi_e \right]^2
      \, \mathrm{d}\Omega.
\end{equation}
In technical terms, this is equivalent to defining the norm in the mis-match
cost function \eqref{eq:mis-match} to be equal to the $H^{-1}_0$ semi-norm. The
$H^{-1}_0$ semi-norm places relatively decreased emphasis on high spatial
wavenumbers, and hence this definition places relatively increased emphasis on
large scale spatial structures in the eddy flux divergence. All results reported
in this article use an $L^2$ norm in the definition of the mis-match function,
although calculations with other norms (not shown) have been performed and are
commented on in the conclusions.


\subsection{Constrained optimisation problem for eddy 
diffusivities}\label{sec:formulation}

Since the force function is defined via the solution of a partial differential
equation, it is natural to redefine the optimisation problem considered at the
end of \S\ref{sec:optimisation} in terms of a partial differential equation
constrained optimisation. For a down-gradient PV parameterisation, letting $V
\subseteq H^1_0 \left( \Omega \right)$ and $V_\kappa \subseteq H^1 \left( \Omega
\right)$ be real Hilbert spaces, a Lagrange constrained cost function
$\hat{\mathcal{J}} : V \times V \times V_\kappa \rightarrow \mathbb{R}$ is
defined, where
\begin{equation}\label{eq:constrained_function}
  \hat{\mathcal{J}}(\Psi_p,\lambda,\kappa) = \left\|\Psi_p - \Psi_e 
\right\|_{L^2}^2
    + \left\langle\grad\lambda, \grad\Psi_p 
      - \kappa\grad\overline{q}\right\rangle_{L^2}
    + \epsilon \mathcal{R} \left( \kappa \right).
\end{equation}
The constrained optimisation problem then seeks a stationary point of the
function $\mathcal{J}(\Psi_p,\lambda,\kappa)$.

The first term in \eqref{eq:constrained_function} is the unconstrained function
\eqref{eq:unconstrained_function}, penalising the mis-match between the
parameterised and diagnosed eddy force functions. The second term is the weak
form partial differential equation constraint. At a stationary point of
$\mathcal{\hat{J}}$, the derivative of $\mathcal{\hat{J}}$ with respect to
$\lambda$ in any direction $\phi \in V$ vanishes, leading to
\begin{equation}\label{eq:weak_form_pde} \left
  \langle \grad \phi, \grad \Psi_p - \kappa \grad \overline{q} \right 
\rangle_{L^2} = \int_\Omega \grad \phi \cdot \left[ \grad\Psi_p - \kappa \grad 
\overline{q} \right] \, \mathrm{d}\Omega = 0 \quad \forall \psi \in V.
\end{equation}
This is a weak form of the Poisson equation \eqref{eq:ffp_poisson}, and hence
$\lambda$ is a Lagrange multiplier enforcing the constraint.

If the final term is absent, then the solution to the constrained optimation
problem finds the optimal $\kappa$ with minimal force function mis-match.
However this problem may be highly ill-conditioned or even ill-posed. The third
term can be used to reguarlise the problem by smoothing the resulting diagnosed
diffusivity at the expense of optimality. A simple form for this regularisation
might, for example, be
\begin{equation}
  \epsilon \mathcal{R} \left( \kappa \right) 
  = \epsilon \left\| \kappa \right\|_{H^1_0}^2
  = \epsilon \int_\Omega \grad \kappa \cdot \grad \kappa\, \mathrm{d}\Omega,
\end{equation}
where $\epsilon \in \mathbb{R}$ is some parameter chosen to control the
smoothness of the resulting optimal $\kappa$.

At a stationary point of the constrained function $\mathcal{\hat{J}}(\Psi_p, 
\lambda, \kappa)$ all derivatives vanish, yielding the optimality system 
\citep[e.g.,][\S 2.2]{Gunzburger-control}
\begin{equation}\label{eq:optimality}
  \ddy{\hat{\mathcal{J}}}{\Psi_p} = 0,\qquad
  \ddy{\hat{\mathcal{J}}}{\lambda} = 0,\qquad
  \ddy{\hat{\mathcal{J}}}{\kappa} = 0,
\end{equation}
where formally the derivatives here are G{\^a}teaux derivatives
\citep[e.g., Ch.~17 of][]{Kantorovich-functional}. This coupled problem can be
solved in its entirety (a ``one-shot'' approach), and where the problem is
non-linear Newton's method can be applied. For cases where the problem is linear
Newton's method formally converges in one iteration. For the applications
considered in this article Newton's method is applied in all cases, and
typically further iterations are applied before tight numerical convergence is
reached. This possibly reflects the ill-conditioned nature of the problems
considered.

A key technical issue encountered here is that the optimality system
\eqref{eq:optimality} changes when components of the constrained function are
modified; this could arise from a switch of mis-match norm, the form of the
parameterisation or the regularisation. If this system is implemented by hand
then the code evaluating the left-hand-side needs to be modified for every
combination of interest. When Newton's method is applied, second derivatives are
required, exacerbating this issue. To bypass the majority of these problems, the
FEniCS automated code generation system is employed \citep{LoggWells10,
Logg-et-al-FENICS, Alnaes-et-al14}, which enables finite element problems to be
described in a high-level syntax and for low-level code to be generated
automatically. In the Python front end, the specification of the cost-function
$\hat{\mathcal{J}}$, its Jacobian and compiling and solving of the optimality
system (via code generation and interfacing with external solver libraries)
translates to the code outlined in Figure~\ref{fig:code}. Different schemes can
be implemented via small code changes: editing \verb!kappa! changes the
definition of the diffusive closure, \verb!J_1! changes the cost function, and
\verb!J_3! changes the regularisation. Although the code may not be as
performant as a hand optimised code, a substantial saving in code development
time easily offsets this, and allows a sweep of a large parameter set that would
have been otherwise be rather inaccessible.

\begin{figure}[tbhp]
  \begin{lstlisting}
    # specify form of kappa
    kappa = xi ** 2

    # form cost function
    J_1 = ((psi - fns["ffd_empb_%i" % (l + 1)]) ** 2) * dx
    J_2_res = grad(psi) - kappa * grad(fns["q_%i_n_mean" % (l + 1)])
    J_2 = inner(grad(lam), J_2_res) * dx
    J_3 = eps_marker * (grad(xi) ** 2) * dx
    J = J_1 + J_2 + J_3

    # compute directional derivative
    dJ = derivative(J, X, du = tests)

    # solve for the system
    solve(dJ == 0, X, boundary_conditions, solver_options)
  \end{lstlisting}
  \caption{Illustrative Python code sample implementing an optimisation problem
  in FEniCS, for the case of PV diffusion with a positive semi-definite
  diffusivity, $\kappa=\xi^2$. The code specifies the desired variant of the
  parameterisation, constructs the cost function ($\mathtt{``ffd\_empb\_\%i"}$
  and $\mathtt{``q\_\%i\_n\_mean"}$ are labels for the diagnosed eddy force
  function associated with the full eddy PV flux and $\overline{q}$), and then
  solves the optimality system.}
  \label{fig:code}
\end{figure}

For all results presented in the article linear systems are solved via
SuperLU and SuperLU\_DIST \citep{Li05, Grigori-et-al07}, via
PETSc \citep[e.g.,][]{Balay-et-al97,petsc-user-ref}.


\section{Diagnostic calculations for the three-layer 
simulation}\label{sect:three_lay}

In this section the eddy force function and mean fields from an eddy resolving
multi-layer QG simulation are used to diagnose eddy diffusivities associated 
with PV and buoyancy mixing parameterisations. The model is described in 
\cite{Maddison-et-al15}. For completeness, the details of the simulation  
are presented here.


\subsection{Simulation details}

The multi-layer QG equations employed here are (e.g.,
\citealt[][\S6.16]{Pedlosky-GFD}; \citealt[][\S5.3.2]{Vallis-GFD})
\begin{equation}
  \ddy{q_i}{t} + \grad\cdot(\ub_{g,i}q_i) = \nu\grad^2 \omega_i
  - r\delta_{in}\omega_i + \delta_{i1}Q_w,
\end{equation}
where the the layer is counted from top (layer $1$) to bottom (layer $n$), the
stream function is defined by $\ub_{g,i} = \hat{\zb}\times\grad\psi_i$, with
$\omega_i = \grad^2 \psi_i$. The layer-wise PV $q_i$ is related to the
stream function $\psi_i$ via
\begin{equation}\begin{aligned}
  q_1 &= \grad^2 \psi_1 + \beta y + s_1^+ (\psi_2 - \psi_1),\\
  q_i &= \grad^2 \psi_i + \beta y 
    + s_i^- (\psi_{i-1} - \psi_i) + s_i^+ (\psi_{i+1} - \psi_i),\\
  q_n &= \grad^2 \psi_n + \beta y + s_n^- (\psi_{n-1} - \psi_n),\\
\end{aligned}\end{equation}
where
\begin{equation}\label{eq:stratification}
  s_i^\pm = \frac{f_0^2}{g_{i\pm1/2} H_i} = 
  \frac{2(f_0^2/N_0^2)_{i\pm1/2}}{(H_i + H_{i\pm1})H_i}
\end{equation}
are stratification parameters, $H_i$ is the thickness of layer $i$, $g_{i+1/2}$
is the reduced gravity at the interface between layers $i$ and $i + 1$, $N_0$ is
the buoyancy frequency, and $f = f_0 + \beta y$ is the Coriolis parameter. The
forcing and dissipation parameters are: a Laplacian viscosity coefficient $\nu$;
a bottom friction coefficient $r$; and the PV tendency due to the wind $Q_w$,
with
\begin{equation}
  Q_w = \begin{cases}
    -\cfrac{\tau_0}{\rho_0}\cfrac{2\pi}{H_1 D} A 
    \sin\left(\pi \cfrac{y_v + D/2}{y_m + D/2}\right), & y_v < y_m\\
    +\cfrac{\tau_0}{\rho_0}\cfrac{2\pi}{H_1 D} \cfrac{1}{A} 
    \sin\left(\pi \cfrac{y_m - y_v}{D/2 - y_m}\right), & \textnormal{otherwise},
  \end{cases}
\end{equation}
where $\tau_0$ is the characteristic magnitude, $\rho_0$ is the reference
density, $x,y\in[0,D]$, $y_v = (y-D/2)$ and $y_m = B(x- D/2)$. The zonal and
meridional directions are $x$ and $y$ respectively. Zero buoyancy boundary
conditions are applied at the top and bottom boundaries \citep{Bretherton66b}. A
partial slip boundary condition $\grad^2 \psi_i = -\alpha^{-1}\grad \psi\cdot
\hat{\nb}$ \citep{Haidvogel-et-al92}, where $\alpha$ is a length scale, is
applied on the lateral boundaries. 

A three-layer, double gyre configuration as detailed in \cite{Marshall-et-al12}
is used. The equations are discretised in space with a conforming triangle
structured mesh with piecewise linear approximation for all fields, a vertex
spacing of $\Delta x = 7.5\ \textrm{km}$, and implemented using the FEniCS
automated code generation system \citep{LoggWells10, Logg-et-al-FENICS,
Alnaes-et-al14}. The model is discretised in time using a third order
Adams--Bashforth scheme with time step size $\Delta t = 20\ \textrm{mins}$,
using the time-stepping approach detailed in \cite{MaddisonFarrell14}. The
equations are integrated for 20,000 days and time averages are taken after this
spin-up period for a further 5,000 days. A summary of the relevant parameters is
given in Table~\ref{tbn:FEMQUOD-param}. For further details about the simulation
set up, see \citet{Marshall-et-al12} and Appendix B of \cite{Maddison-et-al15};
see also \citet{Berloff05a} and \citet{Karabasov-et-al09} for related
configurations of a similar finite difference code on which the finite element
code is based.

\begin{table}[tbhp]
  \begin{center}
  {\small
    \begin{tabular}{|c|c|c|}
      \hline
      parameter & value and units\\
      \hline
      $D$ & $3840\ \mathrm{km}$\\
      $\beta$ & $2\times10^{-11}\ \mathrm{m}^{-1}\ \mathrm{s}^{-1}$\\
      $\tau_0$ & $0.08\ \mathrm{N}\ \mathrm{m}^{-2}$\\
      $\rho_0$ & $1000\ \mathrm{kg}\ \mathrm{m}^{-3}$\\
      $(A, B)$ & $(0.9, 0.2)$\\
      $\nu$ & $100\ \mathrm{m}^2\ \mathrm{s}^{-1}$\\
      $r$ & $4\times10^{-8}\ \mathrm{s}^{-1}$\\
      $\alpha^{-1}$ & $120\ \mathrm{km}$\\
      $(H_1, H_2, H_3)$ & $(0.25, 0.75, 3.00)\ \mathrm{km}$\\
      $(R_1, R_2)$ & $(40, 23)\ \mathrm{km}$\\
      $(s_1^+ H_1 = s_2^- H_2, s_2^+ H_2 = s_3^- H_3)$ & $(2.97, 5.60)\times
      10^{-7}\ \mathrm{m}^{-1}$\\
      $\Delta x$ & $7.5\ \mathrm{km}$\\
      $\Delta t$ & $1200\ \mathrm{s}$ ($=20\ \mathrm{mins}$)\\
      \hline
    \end{tabular}
  }
  \end{center}
  \caption{Summary of simulation parameters used for the three-layer finite
  element ocean gyre calculation, as per \citet{Marshall-et-al12} and
  \citet{Maddison-et-al15}.}
  \label{tbn:FEMQUOD-param}
\end{table}

\begin{figure}[tbhp]
\begin{center}
	\includegraphics[width=0.8\textwidth]{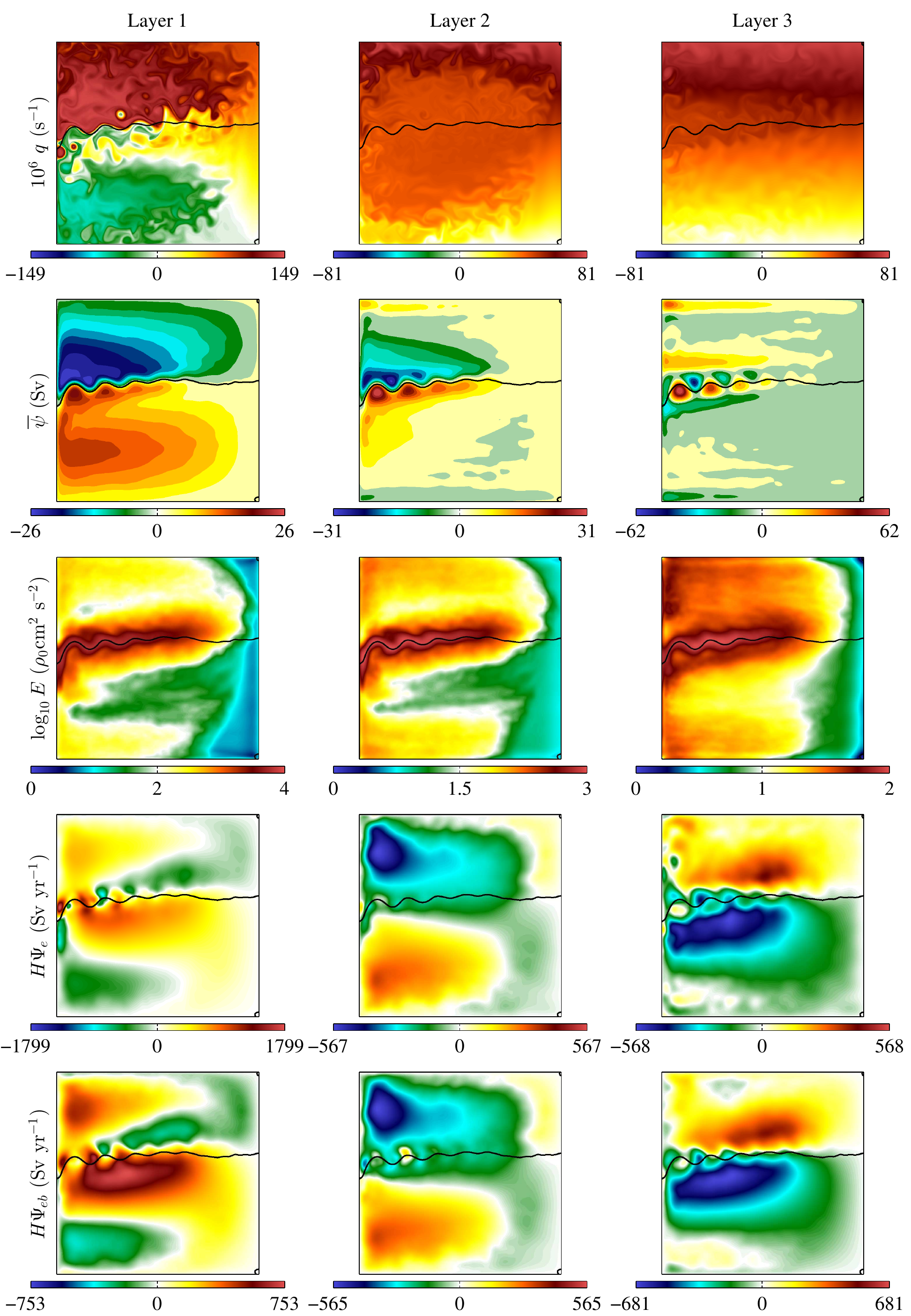}
	\caption{Simulation data for the three-layer finite element ocean gyre
	calculation over the three layers (columns), with (top to bottom row): final
	time PV snapshot (in units of $\mathrm{s}^{-1}$); time-averaged stream
	function $H\overline{\psi}$ (in units of $\mathrm{Sv}$, at 21 contour levels);
	time-averaged total eddy energy $E_i$ (on a logarithmic scale, in units of
	$\rho_0\ \mathrm{cm}^2\ \mathrm{s}^{-2}$); eddy force function $H_i
	\Psi_{e,i}$ from the eddy PV flux (in units of $\mathrm{Sv}\
	\mathrm{yr}^{-1}$); eddy force function $H_i \Psi_{eb,i}$, associated with the
	buoyancy contribution to the eddy PV flux (in units of $\mathrm{Sv}\
	\mathrm{yr}^{-1}$). The black contour is the boundary value of the upper layer
	mean stream function, which approximately indicates the location of the mean
	jet.}
	\label{fig:simulation_data}
\end{center}
\end{figure}

Additional diagnostic quantities were required for the analyis presented here,
absent in the simulation data detailed in \cite{Maddison-et-al15}, and so the
averaging stage was restarted after the 20,000 day spinup. Due to the
sensitive dependence on initial conditions and changes in details such as the
numerical library versions used, the resulting data are not exactly identical to
those presented in \cite{Maddison-et-al15}. Eddy force functions for this
calculation are shown in Figure~\ref{fig:simulation_data}.

\subsection{Eddy diffusivity definition}\label{sect:kappa_defn}

Previously, spatially constant PV diffusivity diagnostics have been reported in
\citet{Maddison-et-al15}; these generally have limited success in minimising the
mis-match between the parameterised and target eddy force function, although
this is not unexpected when making a strong assumption of constant diffusivity. Here
spatially varying diffusivities are considered. Specifically, we consider a:
\begin{itemize}
  \item general case (GEN), a general signed diffusivity $\kappa(\xi) = 
  \xi(\xb)$, supplying no additional information regarding the eddy field and 
  applying no constraints;
  \item positive semi-definite case (POS), where $\kappa(\xi) = \xi^2(\xb) \geq
  0$ excluding the possibility of negative diffusivity. Note that the
  corresponding optimality system is inherently non-linear, and that the zero
  regularisation case is ill-posed (e.g., $\xi\to-\xi$ does not change the value
  of the cost function).
\end{itemize}
Information about the flow may be supplied by, for example, taking 
$\kappa = f(E)\xi$ where $E$ is the eddy energy. Diagnostics of this type will
be discussed in the conclusions.

Considering first diffusion of PV, the layer-wise constrained cost function for
PV diffusion takes the form
\begin{equation}
  \hat{\mathcal{J}}(\Psi_{p,i}, \lambda_i, \xi_i) 
  = \left\|\Psi_{e,i}-\Psi_{p,i}\right\|^2_{L^2} 
    + \left\langle\grad\lambda_i, \grad\Psi_{p,i} -
    \kappa\grad\overline{q}_i\right\rangle_{L^2}
    +\epsilon\mathcal{R}(\xi).
\end{equation}
The resulting optimisation problem for PV diffusion is vertically decoupled and
may be solved layer-wise. The regularisation applied is
\begin{equation}
  \epsilon \mathcal{R} \left( \xi_i\right) 
  = \epsilon \left\| \grad \xi_i \right\|_{H^1_0}^2
  = \epsilon \int_\Omega \grad \xi_i \cdot \grad \xi_i \, \mathrm{d}\Omega.
\end{equation}
For the GEN case this acts to smooth the diffusivity. For the POS case, the
regularisation acts on an auxiliary parameter $\xi_i$ as smoothing the
diffusivity directly would result in an optimisation problem of higher order in
$\xi$, leading to additional numerical difficulties.

In principle the diagnostic may be computed by seeking a value for the
regularisation parameter $\epsilon$ which is as small as possible --- for
example, the value at which the problem becomes sufficiently ill-conditioned for
numerical solver failures to be encountered. Instead a desired spatial scale in
the parameters is chosen here, seeking a value of the regularisation parameter
$\epsilon$ to yield a given spatial ``roughness''. For this, a non-dimensional
roughness measure $\kappa^r$ is defined via an appropriately normalised measure
of the mean square gradient
\begin{equation}
  \kappa^r 
  = D^2\frac{\left\|\kappa\right\|^2_{H^1_0}}{\left\|\kappa\right\|^2_{L^2}}
  = D^2\frac{\int_\Omega \grad \kappa \cdot \grad \kappa\, \mathrm{d}\Omega}
    {\int_\Omega \kappa^2\, \mathrm{d}\Omega}.
\end{equation}
An appropriate value of $\epsilon$ is found via an iterative procedure as
summarised in the pseudo-code in Figure~\ref{fig:pseudocode}. In the majority of
cases, this approach yields a final measured roughness that is within $0.5\%$ of
a target roughness of $\kappa^r = 7500$; for comparison, a field $\kappa =
\sin(20\pi x/D)\sin(20\pi y/D)$ has $\kappa^r = 2(20)^2\pi^2 \approx7900$. In a
minority of cases numerical difficulties mean that small values of $\epsilon$
cannot be reached (due to numerical solver failures), and in these cases the
smallest $\epsilon$ at which convergence is achieved is chosen.

\begin{figure}[tbhp]
  \begin{lstlisting}
    # initialise parameters
    theta = 0.5
    kappa = kappa_init
    eps   = eps_init

    while theta < 0.999:
      # initialise accordingly
      kappa = solve(..., kappa_init, eps) 
      
      # decrease epsilon and continue
      if roughness < tolerance: 
        eps = theta * eps
        kappa_init = kappa
      # if tolerance exceeded, try again with larger eps initialised 
      # with previous kappa
      elif:
        theta = theta * 4/3
        eps = theta * eps
        
  \end{lstlisting}
  \caption{Pseudo-code for the procedure employed to select a value of the
  regularisation parameter $\epsilon$ so as to yield a parameter with a given
  degree of ``roughness''. An initial $\epsilon$ is chosen and decreased
  geometrically by some factor $\theta < 1$, until non-convergence or the
  roughness condition is triggered. The loop is reinitialised at the previously
  converged solution, with the value of the factor $\theta$ increased. This
  continues until some tolerance for $\theta$ is triggered.}
  \label{fig:pseudocode}
\end{figure}

To quantify the diagnosed diffusivity, six measures are utilised: (i) the mean
diffusivity; (ii) an eddy energy weighted mean diffusivity; (iii) a diffusivity
positivity measure; (iv) the $L^2$ correlation between the diffusivity and the
eddy energy; (v) the roughness of the diffusivity; (vi) the $L^2$ relative
mis-match error between the parameterised and target eddy force function. The
mean diffusivity and eddy energy weighted mean diffusivity, are defined via
\begin{equation}\label{eq:kappa_mean}
  \kappa^{m} = \frac{\int_\Omega \kappa\, \mathrm{d}\Omega}
    {D^2},\qquad
  \kappa^{m}_E = \frac{\int_\Omega E\kappa\, \mathrm{d}\Omega}
    {\int_\Omega E\, \mathrm{d}\Omega}.
\end{equation}
Positivity is measured via
\begin{equation}\label{eq:kappa_pos}
  \kappa^{>0} = \frac{\int_\Omega \mathcal{H}(\kappa)\, \mathrm{d}\Omega}
    {D^2},
\end{equation}
where $\mathcal{H}(\kappa)$ is the Heaviside function, equal to one where
$\kappa \geq 0$ and zero otherwise. The correlation between $\kappa$ and the
eddy energy is measured via
\begin{equation}\label{eq:correl}
  \mbox{corr}\left(\kappa, E  \right) 
  = \frac{\langle \kappa, E \rangle_{L^2}}{\|\kappa\|_{L^2} \|E\|_{L^2}}
  = \frac{\int_\Omega \kappa E\, \mathrm{d}\Omega}
    {\sqrt{\int_\Omega \kappa^2\, \mathrm{d}\Omega}\sqrt{\int_\Omega E^2\, 
\mathrm{d}\Omega}}.
\end{equation}
Note that the correlation is bounded, $-1 \leq \mbox{corr}\left(\kappa, E
\right) \leq 1$. The (non-dimensional) roughness of the diffusivity is measured
via
\begin{equation}\label{eq:kappa_rough}
  \kappa^{r}
  =  D^2
    \frac{\left\|\kappa\right\|^2_{H_0^1}}{\left\|\kappa\right\|^2_{L^2}}
  = D^2
    \frac{\int_\Omega \grad \kappa \cdot \grad \kappa\, \mathrm{d}\Omega}
    {\int_\Omega \kappa^2\, \mathrm{d}\Omega}.
\end{equation}
Finally mis-match between parameterised and diagnosed force functions is
measured via an $L^2$ relative error
\begin{equation}\label{eq:L2_err}
  \mathcal{E}_{L^2} 
  = \frac{\left\|\Psi_e - \Psi_p\right\|_{L^2}}{\left\|\Psi_e\right\|_{L^2}}
  = \sqrt{\frac{\int_\Omega \left(\Psi_e - \Psi_p\right)^2\, \mathrm{d}\Omega}
    {\int_\Omega \Psi_e^2\, \mathrm{d}\Omega}}.
\end{equation}


\subsection{Results: PV diffusion}

\begin{figure}[tbhp]
\begin{center}
\includegraphics[width=0.75\textwidth]
  {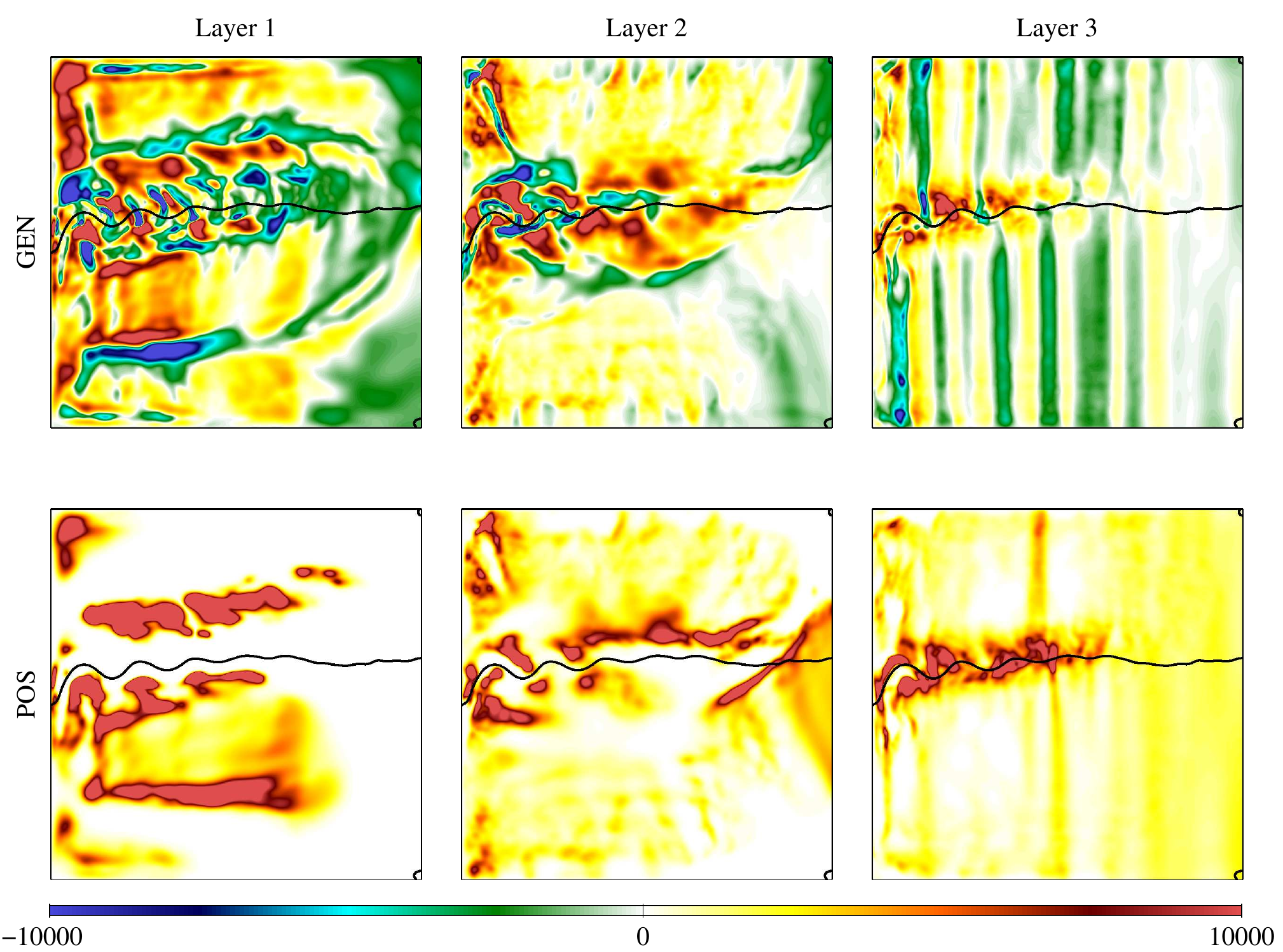}
	\caption{The diffusivity $\kappa$ (in units of $\mathrm{m}^2\
	\mathrm{s}^{-1}$) associated with PV diffusion over the three layers
	(columns), for the GEN case $\kappa = \xi$ (top row) and POS case $\kappa =
	\xi^2\geq 0$ (bottom row). The colour scale is fixed and saturated.}
	\label{fig:kappa_PV_vary_variety_L2}
\end{center}
\end{figure}

\begin{figure}[tbhp]
\begin{center}
	\includegraphics[width=0.75\textwidth]{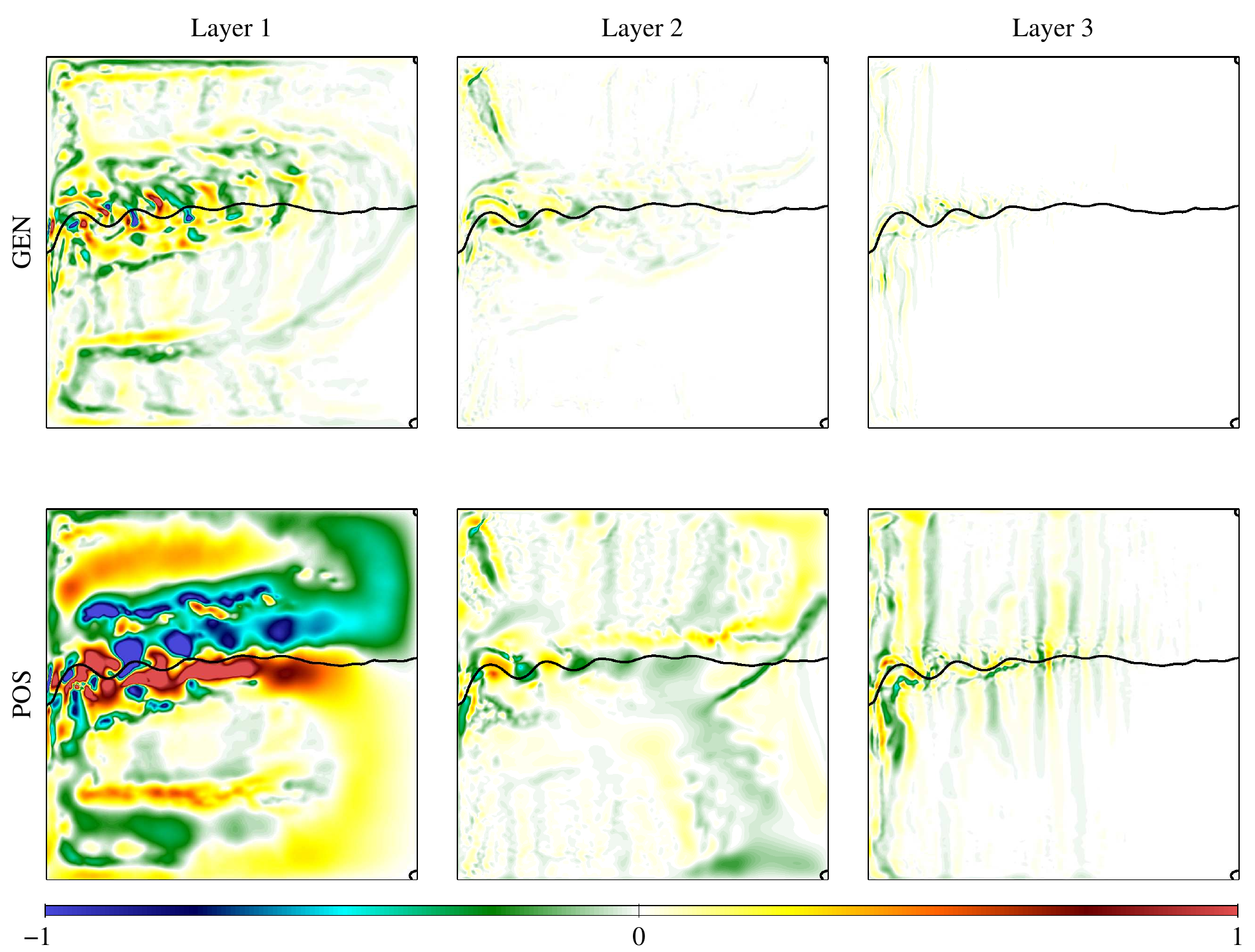}
	\caption{Layer-wise non-dimensional mis-match $D \left(\Psi_{e,i} - \Psi_{p,i}
	\right) / \left\| \Psi_{e,i}\right\|_{L^2}$ associated with PV diffusion over
	the three layers (columns), for the GEN case $\kappa = \xi$ (top row) and POS
	case $\kappa = \xi^2\geq 0$ (bottom row). The colour scale is fixed and
	saturated in layer 1.}
	\label{fig:kappa_PV_L2_raw_error}
\end{center}
\end{figure}

\begin{figure}[tbhp]
\begin{center}
  \includegraphics[width=0.9\textwidth]{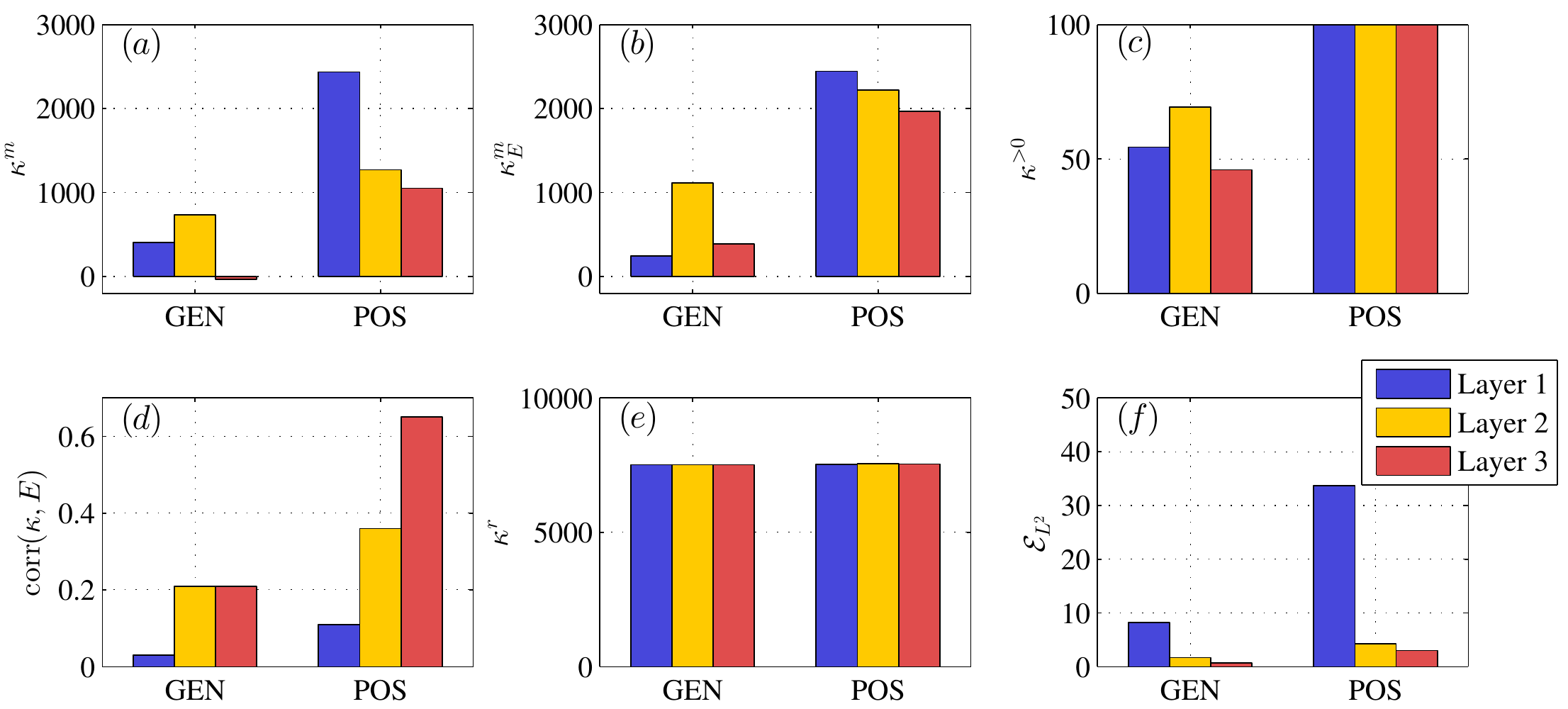}
  \caption{Bar graphs comparing the diagnostic data across the three layers of
  for the GEN and POS associated with PV diffusion : ($a,b$) the mean $\kappa^m$
  and the eddy energy weighted mean $\kappa^m_E$ from equation
  \eqref{eq:kappa_mean}; ($c$) the positivity index $\kappa^{>0}$ from equation
  \eqref{eq:kappa_pos}; ($d$) the correlation $\mbox{corr}(\kappa, E)$ from
  equation \eqref{eq:correl}; ($e$) the roughness $\kappa^r$ from equation
  \eqref{eq:kappa_rough}; ($f$) the relative $L^2$ error $\mathcal{E}_{L^2}$
  from equation \eqref{eq:L2_err}.}
  \label{fig:kappa_pv_L2_bar}
\end{center}
\end{figure}

Informed by resolution tests, the parameterised force function and parameter
$\xi$ are computed on a structured conforming triangle mesh with nodal spacing
$\Delta x = 15\ \textrm{km}$ for all cases presented in the following sections.
The diagnosed model force function $\Psi_{e,i}$ from the finite element
simulation at resolution $\Delta x = 7.5\ \textrm{km}$ is interpolated onto this
coarser resolution grid via consistent interpolation (evaluation of the higher
resolution data at the vertices of the coarse grid).
Figure~\ref{fig:kappa_PV_vary_variety_L2} shows the diffusivity $\kappa$
diagnosed for the GEN and POS diffusivity variants. The local mis-match is shown
in Figure~\ref{fig:kappa_PV_L2_raw_error}. Values for the diagnostic quantities
from equation \eqref{eq:kappa_mean} to \eqref{eq:L2_err} are summarised in
Figure~\ref{fig:kappa_pv_L2_bar}.

Starting with the GEN case, there are regions of negative diffusivity towards
the eastern boundary in the upper and middle layers, and a correspondingly large
positive diffusivity towards the western boundary, at least in the upper layer.
This is consistent with the signal that might be associated with a westward
propagating of eddy activity. There is a second pool of negative diffusivity
towards the down-stream mean jet in the upper layer. This is consistent with an
outward flux of activity due to the ``wave radiator'' mechanism discussed in
\citet{WatermanJayne11, WatermanJayne12} for the stable down-stream region of an
inertial barotropic jet. In the middle layer, a comparison with the mean
streamlines $\overline{\psi}$ in Figure~\ref{fig:simulation_data} (second row)
reveals that the closed streamlines north and south of the jet correlate with
regions of positive diffusivity. This is in agreement with the principle of PV
homogenisation within closed streamlines \citep{RhinesYoung82}. A similar
correlation exists in the upper layer, though this is less strong; this
correlation breaks down to the north of the mean jet, possibly due to the
presence of strong wind forcing in this layer. There are signals of negative
diffusivity in the upper layer confined close to the northern and southern
boundaries. This signal can be expected if there is a local eddy activity
backscatter owing to the presence of Fofonoff gyres in these regions
\cite[e.g.,][]{Berloff05b, MarshallAdcroft10}. In the lower layer, the
diffusivity is large and positive in the jet, correlating with the location of
the largest eddy energy. However in this layer there are meridionally oriented
patterns in the diffusivity away from the jet. This ``banding'' correlates with
a similar pattern in the local mis-match in
Figure~\ref{fig:kappa_PV_L2_raw_error}, and so it is possible that this signal
is a numerical artefact. A similar effect may account for the alternating
diffusivity sign in the upper layer mean jet. It is apparent that there are
regions of significant negative diffusivity.

For the GEN case, in the middle layer unweighted and eddy energy weighted means
are positive and of a similar magnitude (around $750$ and $1200\ \mathrm{m}^2\
\mathrm{s}^{-1}$ respectively), indicating that the diffusivity is largely
positive in this layer. This is supported by the positivity index in the middle
layer at around $60\%$. Some degree of correlation between eddy energy and
diffusivity is seen. The roughness is found to be well controlled by the
solution output criterion; it has been confirmed that the $\kappa$ output is
within $0.5\%$ of the fixed target roughness. Given this, we see that the
resulting $L^2$ relative error is low, at less than $2\%$. In the lower layer
the eddy energy weighted mean is smaller, and the unweighted mean in negative.
While the error in the inversion is well controlled, at less than $1\%$, the
banding of positive and negative diffusivities away from the jet in this case
lead to a negative unweighted mean. This may reflect difficulties in the
diagnostic in this region. In the upper layer the diffusivity, while positive in
the means, exhibits almost no correlation with the eddy energy and, for a given
roughness, the relative $L^2$ mis-match is greater than in the other two layers.

The POS case shows similar patterns of positive diffusivity around the location
of the mean jet and towards the western boundary. However, this diagnostic shows
large regions of very low diffusivity, which typically correlate with regions of
negative diffusivity seen in the GEN case. In the middle and lower layer the
correlation between diffusivity and eddy energy has increased compared to the
GEN case. The corresponding relative $L^2$ mis-match is slightly larger than the
GEN case, by at less than $5\%$ for a similar level of roughness. In the upper
layer the $L^2$ mismatch is much larger, at around $30\%$, and a very low
correlation between diffusivity and eddy energy is observed. On closer
inspection of the spatial distribution of error, seen in
Figure~\ref{fig:kappa_PV_L2_raw_error}, the errors are generally large around
the mean jet. This is particularly the case in the upper layer.

In summary, the diagnostic calculations produce a diffusivity field that
correlates with some physical processes that are known to occur. In the middle
and lower layer, both diffusivity variants shows a strong positive signal that
has some correlation with the eddy energy and, for a given roughness, the
resulting $L^2$ mis-match is low. The same cannot be said for the diganosed
diffusivity in the upper layer, where the correlation between eddy energy and
diffusivity is low, and the errors are significantly larger for a given
roughness. It appears that a negative signal is a prevalent especially in the
upper layer; for a given roughness, the POS case has associated with it
significantly larger mis-match error.


\subsection{Results: Buoyancy mixing}

An analogous procedure may be applied to the down-gradient mixing of buoyancy
\begin{equation}\label{eq:GM-param}
  \overline{\ub'b'} = -\kappaGM\grad\overline{b},
\end{equation}
which, in the QG setting, is equivalent to the Gent--McWilliams (GM)
parameterisation (\citealt{GentMcWilliams90}; see also
\citealt{Treguier-et-al97}). The analogous constrained cost function in the
continuously stratified setting is given by
\begin{equation}\label{s4:J-cont}
  \hat{\mathcal{J}}\left(\Psi_p, \lambda, \kappaGM\right) 
    = \int_{z=-H}^0 \left[\left\|\Psi_{eb}-\Psi_p\right\|_{L^2}^2 +
    \left\langle\grad\lambda, \grad\Psi_p 
      - \ddy{}{z}\left(\frac{f_0}{N_0^2}\kappaGM\grad\overline{b}\right)
      \right\rangle_{L^2}
    +\epsilon\left\|\grad\kappaGM\right\|_{L^2}^2\right] \, \mathrm{d}z,
\end{equation}
where all inner products and norms are defined via integration over the
horizontal domain. The eddy force function associated with the buoyancy fluxes
is shown in the lower row of Figure~\ref{fig:simulation_data}.

In the multi-layer quasi-geostrophic equations the buoyancy flux and
$\grad\overline{b}$ are defined on interfaces and so, via equation
\eqref{eq:GM-param}, $\kappaGM$ is also interfacial. The corresponding PV flux
is related to the interfacial buoyancy flux via a vertical derivative operator
\citep{GreatbatchLamb90}. This introduces vertical coupling, and as such the
corresponding optimisation problem for $\kappaGM$ is fully three-dimensional,
unlike the previous PV diffusion case. An alternative method, not pursued here,
is to define an interfacial eddy stress function \citep[][Appendix
C]{Maddison-et-al15}, and use this as the basis for an $\kappaGM$ diagnostic
computed separately on each interface.

The eddy buoyancy fluxes on each interface $(R,S) = (f_0^2/N_0^2)
\overline{\ub'(\dy\psi'/\dy z)}$ may be defined
\begin{equation}
  R_{i+1/2} = -\frac{1}{2}\left(\ddy{}{y}(\psi_i + \psi_{i+1})\right)
    H_i s_i^+ (\psi_i - \psi_{i+1}),\qquad
  S_{i+1/2} = +\frac{1}{2}\left(\ddy{}{x}(\psi_i + \psi_{i+1})\right)
    H_i s_i^+ (\psi_i - \psi_{i+1}),
\end{equation}
with stratification parameters $s_i^+$ as given by \eqref{eq:stratification}.
Vertical differencing then leads to a discrete eddy PV flux associated with the
eddy buoyancy fluxes (i.e. these are the vertical stresses appearing in a
vertically discrete Taylor--Bretherton identity \citep[][appendix
B]{Maddison-et-al15}. The interfacial GM coefficient is then defined via
\begin{equation}
  \left(R_{i+1/2}, S_{i+1/2}\right) = 
    H_i s_i^+ \left(-(\kappaGM)_{i+1/2} 
    \grad\left(\overline{\psi}_i - \overline{\psi}_{i+1}\right)\right).
\end{equation}
Again, the GEN case $\kappaGM = \xi$ and POS case $\kappaGM = \xi^2$ are
considered. The vertically discrete cost function for the buoyancy mixing 
case is given by
\begin{equation}\label{s4:J}
  \hat{\mathcal{J}} = \sum_{i=1}^3 \left\|H_i(\Psi_{eb,i}-\Psi_{p,i})\right\|^2_{L^2} 
    + \sum_{i=1}^{2}\left(\left\langle\grad\lambda_i, \grad\Psi_{p,i} +
      H_i s_i^+ (\kappaGM)_{i+1/2}
      (\overline{\psi}_i - \overline{\psi}_{i+1})\right\rangle_{L^2}
    +\epsilon\frac{H_i+H_{i+1}}{2}
    \left\|\grad\xi_{i+1/2}\right\|_{L^2}^2
    \right).
\end{equation}
The regularisation again penalises gradients in $\kappaGM$, but without
increasing the order for the resulting optimisation problem for the POS case.
The procedure for implementation, solving the variational problem, simulation
details, manner of decreasing $\epsilon$ and output of solution based on the
roughness criteria (with target roughness of 7500 as for the PV diffusion case)
are as detailed in the previous subsection, where the roughness is now defined
to be
\begin{equation}\label{eq:GMroughness}
  \kappa^r_{\scriptsize\mbox{gm}} = 
    D^2 \cfrac{\mathlarger{\sum\limits_{i=1}^2}\left(\cfrac{H_i + H_{i+1}}{2} 
      \left\|\grad\kappa_{\scriptsize\mbox{gm},\ i+1/2}\right\|_{L^2}^2\right)}{
    \mathlarger{\sum\limits_{i=1}^2}\left(\cfrac{H_i + H_{i+1}}{2} 
      \left\|\kappa_{\scriptsize\mbox{gm},\ i+1/2}\right\|_{L^2}^2\right)}.
\end{equation}

\begin{figure}[tbhp]
\begin{center}
	
\includegraphics[width=0.8\textwidth]
  {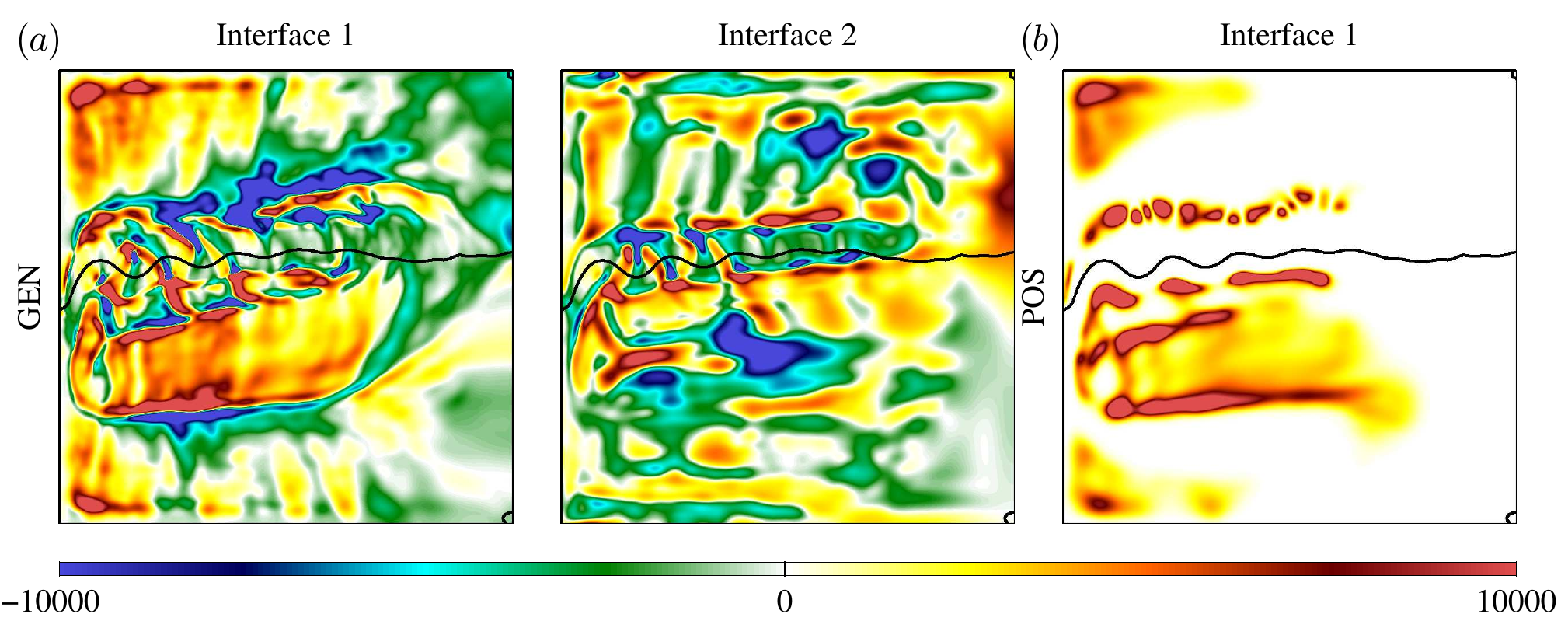}
	\caption{The diffusivity $\kappaGM$ (with units of $\mathrm{m}^2\
	\mathrm{s}^{-1}$) on the interfaces associated with buoyancy mixing. ($a$) 
  the GEN case $\kappaGM = \xi$ for both interfaces; ($b$) the POS case with 
  $\kappaGM = \xi^2$ for upper interface only, as the lower interface is the 
  zero solution. The colour scale is fixed and saturated.}
	\label{fig:kappa_gm_vary_variety_L2}
\end{center}
\end{figure}

\begin{figure}[tbhp]
\begin{center}
	\includegraphics[width=0.75\textwidth]{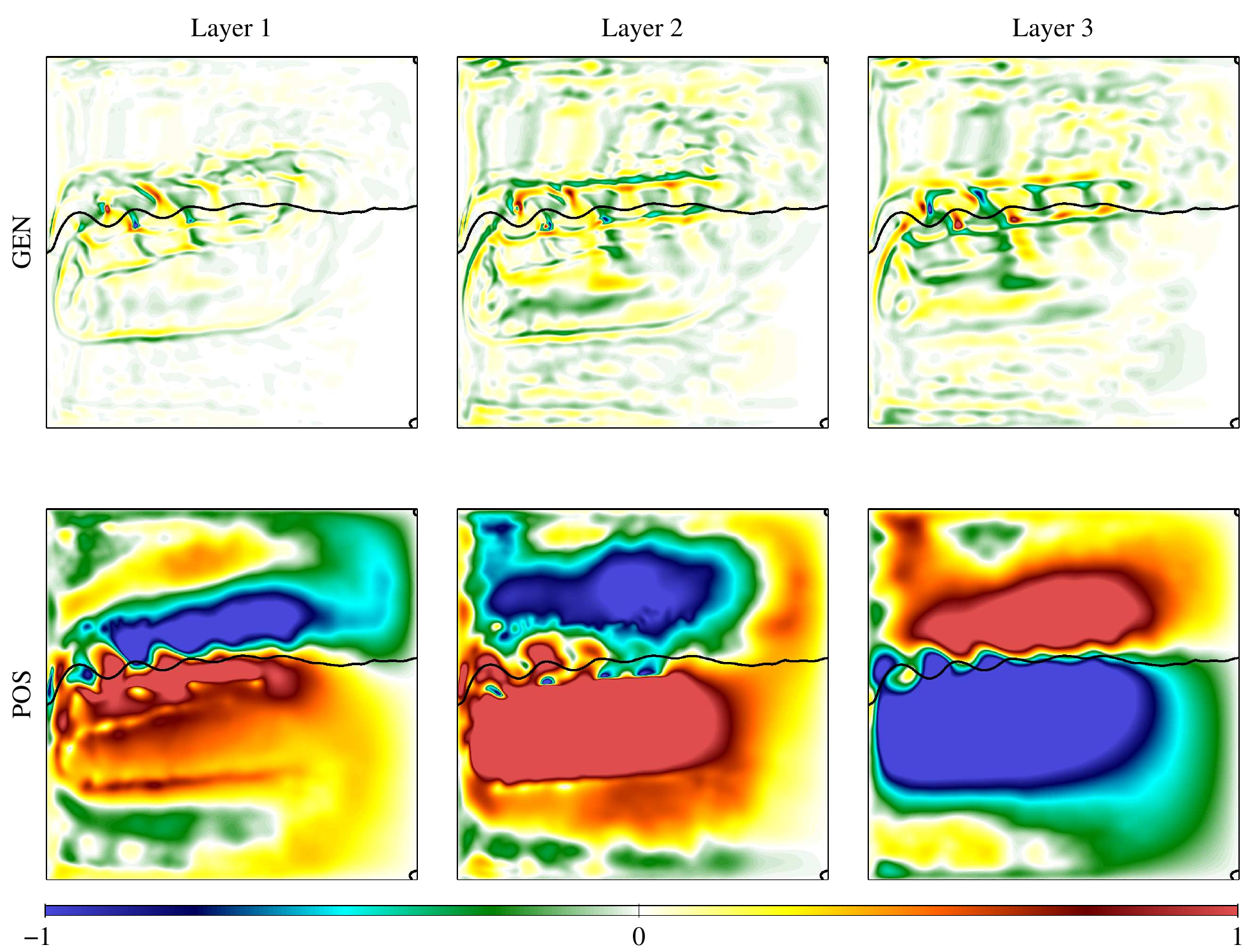}
  \caption{Layer-wise non-dimensional mis-match $D \left(\Psi_{eb,i} -
  \Psi_{p,i} \right) / \left\| \Psi_{eb,i}\right\|_{L^2}$ associated with
  buoyancy mixing over the three layers (columns), for the GEN case $\kappaGM =
  \xi$ (top row) and the POS case $\kappaGM = \xi^2 \geq 0$ (bottom row). The
  colour scales are fixed and saturated in the upper and bottom layer for the
  GEN case and across all layers in the POS case.}
  \label{fig:kappa_gm_raw_error}
\end{center}
\end{figure}

\begin{figure}[tbhp]
\begin{center}
	\includegraphics[width=0.9\textwidth]{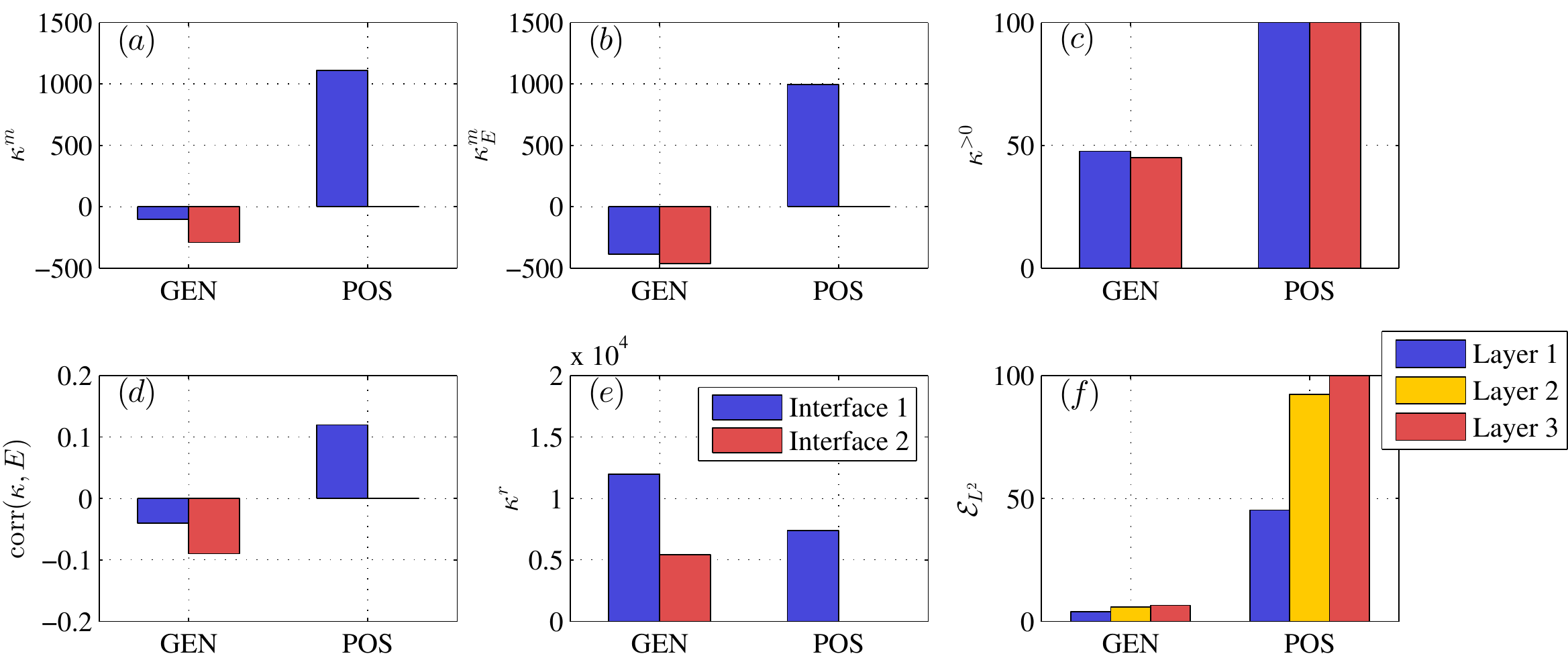}
	\caption{Bar graph comparing the diagnostic data across the two interfaces and
	three layers of the GEN and POS case for buoyancy mixing: ($a,b$) the mean
	$\kappa^m$ and the eddy energy weighted mean $\kappa^m_E$ from equation
	\eqref{eq:kappa_mean}; ($c,d$) the positivity index $\kappa^{>0}$ and
	eddy energy weighted positivity index $\kappa^{>0}_E$ from equation
	\eqref{eq:kappa_pos}; ($e$) the roughness $\kappa^r$ from equation
	\eqref{eq:kappa_rough}; ($f$) the relative $L^2$ error $\mathcal{E}_{L^2}$
	from equation \eqref{eq:L2_err}. Note that the lower interface solution for
	the POS case is zero.}
	\label{fig:kappa_gm_L2_bar}
\end{center}
\end{figure}

The resulting interfacial GM coefficients are shown in
Figure~\ref{fig:kappa_gm_vary_variety_L2}, and the local mis-matches are shown
in Figure~\ref{fig:kappa_gm_raw_error}. The same diagnostic quantities from
equation \eqref{eq:kappa_mean} to \eqref{eq:L2_err} are employed to assess the
resulting diffusivity, and these are summarised in
Figure~\ref{fig:kappa_gm_L2_bar}.

Consider first the GEN case, shown in
Figure~\ref{fig:kappa_gm_vary_variety_L2}($a$). In the upper interface
$\kappaGM$ is positive in the north-west and south-west corners. A large region
of positive $\kappaGM$ exists in the southern gyre. There is a significant
pattern of negative $\kappaGM$, particularly to the north of the mean jet and in
the downstream mean jet. In the lower interface, $\kappaGM$ is predominantly
negative around the down-stream mean jet. This negative coefficient is
consistent with previously reported signals of baroclinic stability here,
described in \cite{Berloff05a} and \cite{Maddison-et-al15}. Away from the jet
there is a positive $\kappaGM$ region towards the north-east, but negative
$\kappaGM$ in the southern gyre. The unweighted and the eddy energy weighted
means are negative, especially in the lower interface. The positivity index is
low, below 50\%, and the correlation between $\kappaGM$ and the eddy energy is
low and negative, indicating the prevalence of a negative signal and a weak
correlation with eddy energy. The $L^2$ relative errors however are reasonable,
at less than $10\%$ for both interfaces. An observation to be made here is that,
unlike the PV diffusion case, here the upper layer has the lowest mis-match.
That is, the use of a global mis-match cost function here has led to a
preferential decrease in the upper layer mis-match, at the expense of the lower
two layers.

Now considering the POS case, the lower layer diffusivity is zero (not shown).
This was found to be robust even after, in the algorithm of
Figure~\ref{fig:pseudocode}, choosing multiple initial values of $\epsilon$ and
multiple initial guesses for the $\xi$ field. The existence of a global minimum
with non-zero lower interface $\kappaGM$ cannot be ruled out. In the upper
interface, however, a non-zero solution is found, with a strong positive signal
in the southern gyre and towards the north-western and south-western boundaries.
Here, regions of low diffusivity correlate well with the regions of negative
diffusivity previously observed in the GEN case. The associated error is large
almost everywhere, as seen in Figure~\ref{fig:kappa_gm_raw_error} and
Figure~\ref{fig:kappa_gm_L2_bar}($f$).

In summary, the GEN case diagnosed diffusivity shows strong negative signals,
for example in the lower interface down-stream mean jet. Enforcing positive
semi-definite diffusivity in the POS case leads to very significantly increased
mis-match errors, and difficulty in diagnosing a non-trivial diffusivity in the
lower interface. The correlation with eddy energy is, in both cases, low.

A key issue encountered here is that, in a three-layer configuration, each of
the two interfaces is coupled to layers which experience either direct wind
forcing or bottom dissipation. Hence more significant influence from forcing and
dissipation may be expected in these diagnostics. This is addressed in the
following section by adding an increased number of model layers.


\section{Results: Five layers, potential vorticity and buoyancy mixing}

A five layer simulation is performed using a higher horizontal resolution model
with a grid spacing of $\Delta x = 3.25\ \textrm{km}$, using a finite difference
code (see e.g., \citet{Berloff05a}, and particularly \citet{Karabasov-et-al09}
for the CABARET numerical scheme which is used here). Parameter values that
differ from the earlier three-layer calculation are given in
Table~\ref{tbn:PEQUOD-param}. Stratification parameters are based upon
stratification profiles from the World Ocean Circulation Experiment
\citep{GouretskiKolterman04, Kolterman-et-al11} data, employing a density
profile of the form $\rho(z) = a + b\ex^{z/c}$ (noting that $z=0$ is the top of
the ocean); specific values of $a$, $b$ and $c$ as well as $f_0$ are also given
in Table~\ref{tbn:PEQUOD-param}. Note that the leading baroclinic deformation
radii are somewhat lower than the earlier three-layer calculation.

\begin{table}[tbhp]
  \begin{center}
  {\small
    \begin{tabular}{|c|c|c|}
      \hline
      parameter & value and units\\
      \hline
      $\nu$ & $10\ \mathrm{m}^2\ \mathrm{s}^{-1}$\\
      $(H_1, H_2, H_3, H_4, H_5)$ & $(0.15, 0.29, 0.58, 1.16, 2.32)
        \ \mathrm{km}$\\
      $(R_1, R_2, R_3, R_4)$ & $(33, 17, 11, 10)\ \mathrm{km}$\\
      $s_1^+ H_1 = s_2^- H_2$ & $ 8.09 \times 10^{-7}\ \mathrm{m}^{-1}$\\
      $s_2^+ H_2 = s_3^- H_3$ & $ 7.24 \times 10^{-7}\ \mathrm{m}^{-1}$\\
      $s_3^+ H_3 = s_4^- H_4$ & $ 1.16 \times 10^{-6}\ \mathrm{m}^{-1}$\\
      $s_4^+ H_4 = s_5^- H_5$ & $ 5.90 \times 10^{-6}\ \mathrm{m}^{-1}$\\
      $\Delta x$ & $3.25\ \mathrm{km}$\\
      $\Delta t$ & variable, based on the Courant number\\
      \hline
      $a$ & $1000 \ \mathrm{kg}\ \mathrm{m}^{-3}$\\
      $b$ & $1.2 \ \mathrm{kg}\ \mathrm{m}^{-3}$\\
      $c$ & $500 \ \mathrm{m}$\\
      $f_0$ & $\cfrac{2\pi}{3600\times24}
        \sin\left(\cfrac{50^{\circ}\pi}{180^{\circ}}\right)
        \ \mathrm{s}^{-1}$\\
      \hline
    \end{tabular}
  }
  \end{center}
  \caption{Simulation parameters used for the five-layer finite difference ocean
  gyre calculation. Other parameters employed are as per
  Table~\ref{tbn:FEMQUOD-param}.}
  \label{tbn:PEQUOD-param}
\end{table}

Diagnostic calculations are repeated for this case via interpolation of the
finite difference data onto the earlier finite element mesh with a nodal spacing
of $\Delta x = 15\ \mathrm{km}$. Diffusivities are then diagnosed as before.


\subsection{Potential vorticity diffusion}

\begin{figure}[tbhp]
\begin{center}
  \includegraphics[width=\textwidth]
    {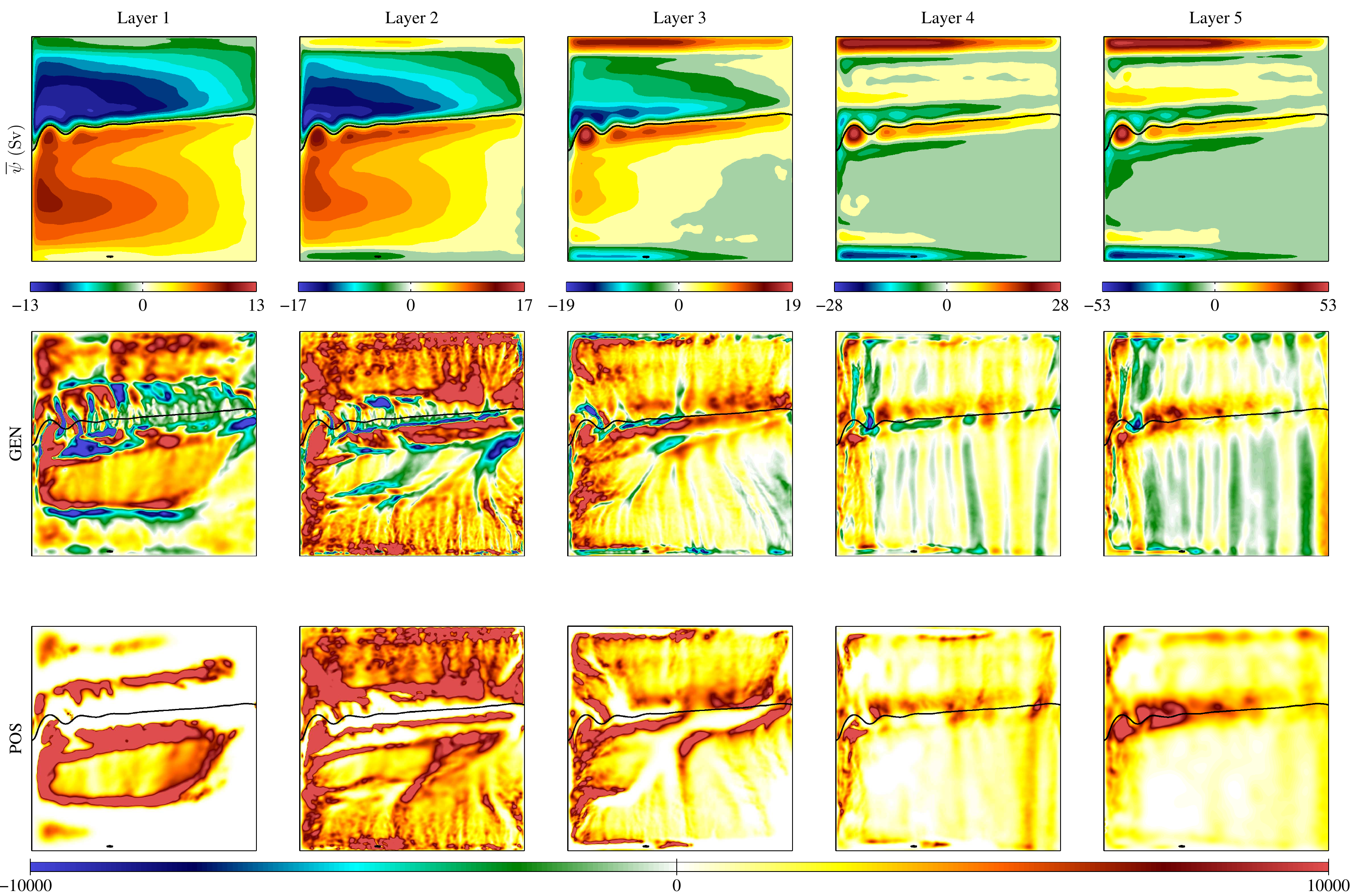}
  \caption{Contours of $\overline{\psi}$ (with units of $\mathrm{Sv}$) at 21
  contour levels (top row) and the diffusivity $\kappa$ (with units of
  $\mathrm{m^2}\ \mathrm{s}^{-1}$) associated with the GEN case $\kappa = \xi$
  (middle row) and POS case $\kappa = \xi^2 \geq 0$ (bottom row) for PV
  diffusion over the five layers (columns). The colour scale for the diffusivity
  is saturated.}
  \label{fig:kappa_archer_PV_L2}
\end{center}
\end{figure}

\begin{figure}[tbhp]
\begin{center}
  \includegraphics[width=0.9\textwidth]{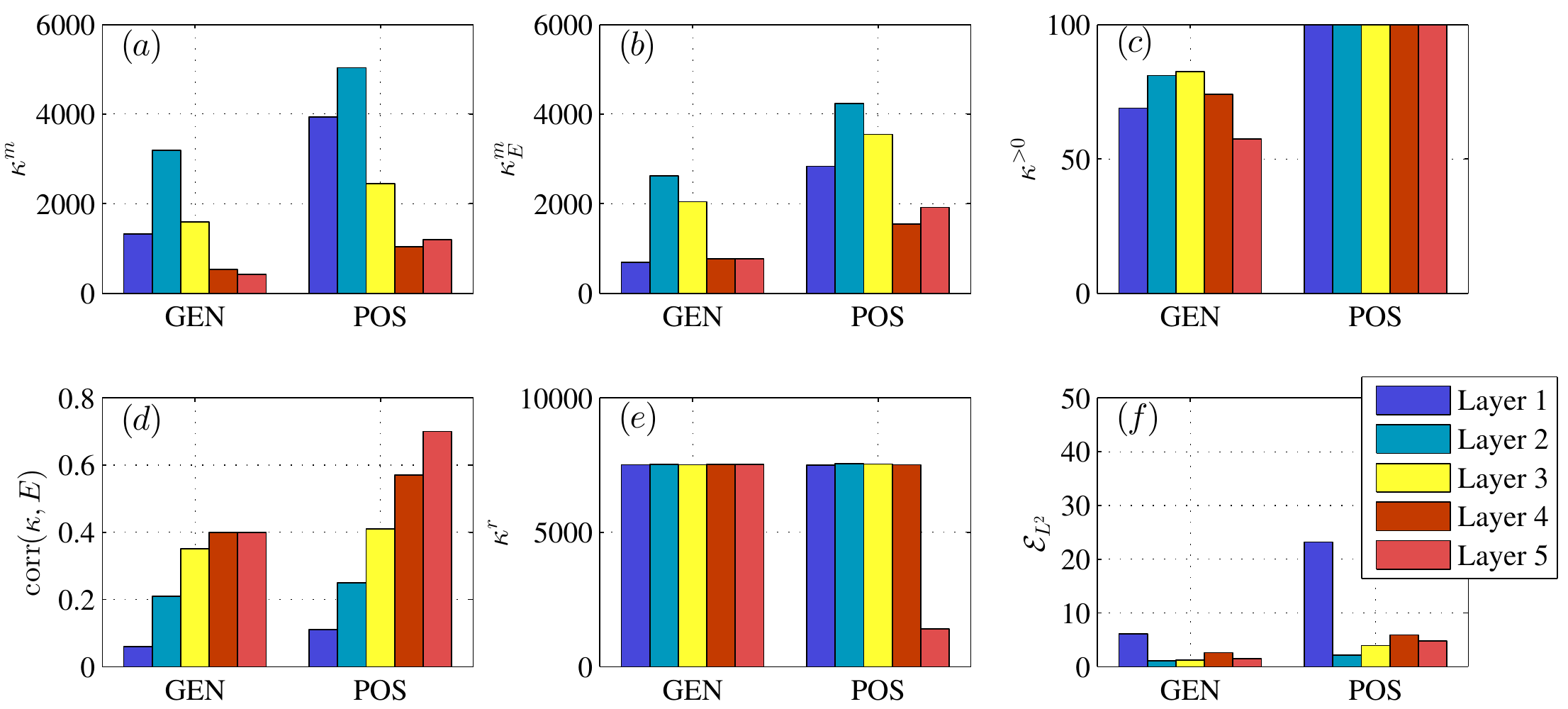}
  \caption{Bar graph comparing the diagnostic data across the five layers of the
  GEN and POS case associated with down-gradient PV diffusion: ($a,b$) the mean
  $\kappa^m$ and the eddy energy weighted mean $\kappa^m_E$ from equation
  \eqref{eq:kappa_mean}; ($c,d$) the positivity index $\kappa^{>0}$ and eddy
  energy weighted positivity index $\kappa^{>0}_E$ from equation
  \eqref{eq:kappa_pos}; ($e$) the roughness $\kappa^r$ from equation
  \eqref{eq:kappa_rough}; ($f$) the relative $L^2$ error $\mathcal{E}_{L^2}$
  from equation \eqref{eq:L2_err}. The lowest layer for the POS case has not
  been returned via triggering the roughness criterion, and instead the last
  converged solution has been returned.}
  \label{fig:kappa_pv_archer_L2_bar}
\end{center}
\end{figure}

The corresponding eddy force function $\Psi_e$ and total eddy energy
distribution $E$ for the five-layer calculation are largely similar in structure
to the three layer case shown in Figure~\ref{fig:simulation_data}. As a
consequence of this the resulting PV diffusivities associated with the GEN and
POS cases, displayed in Figure~\ref{fig:kappa_archer_PV_L2}, show largely
similar structures to the three-layer case. The associated diagnostic quantities
from equation \eqref{eq:kappa_mean} to \eqref{eq:L2_err} are summarised in
Figure~\ref{fig:kappa_pv_archer_L2_bar}.

Considering first the GEN case, the resulting diffusivity is predominantly
positive over all layers, though still possessing significant local negative
signals particularly in the upper layer. The rapidly varying structure within
the mean jet is present but may again be seen to be correlated with the
locations of largest local error (not shown; cf.
Figure~\ref{fig:kappa_PV_L2_raw_error}). In layer three there are suggestions of
a boundary confined negative signal near the north and southern boundaries.
There is also a suggestion of a negative signal around the mean jet in the
second and third layers. There is again correlation between locations of
positive diffusivity within contours of closed stream lines $\overline{\psi}$,
especially in the third and fourth layers. The overall positivity for the
diagnosed diffusivity is generally high (see
Figure~\ref{fig:kappa_pv_archer_L2_bar}$a,b,c$). Further, there is mild
correlation between the diffusivity and the eddy energy and, for the same given
roughness as in the three layer case, the resulting $L^2$ mis-match errors are
all less than $10\%$. The mis-match is particularly low away from the upper
layer.

For the POS case, the observations are again similar to those made for the three
layer case. The regions of positive diffusivity in the GEN and POS case largely
coincide, with strong positive diffusivity in the southern gyre and western
boundary. These regions of positive diffusivity again correlate well with the
locations where there are closed mean stream lines. Regions of low diffusivity
also correlate well with the regions of negative diffusivity present in the GEN
case. There is again evidence of correlation between the diffusivity and the
eddy energy especially in the lower layers and, for a given roughness, the $L^2$
mis-match is reasonable away from the upper layer. It should be noted that the
bottom layer solution for the POS case has not converged to the target
roughness, though the errors are still less than $10\%$.


\subsection{Buoyancy mixing}

\begin{figure}[tbhp]
\begin{center}
	
\includegraphics[width=\textwidth]{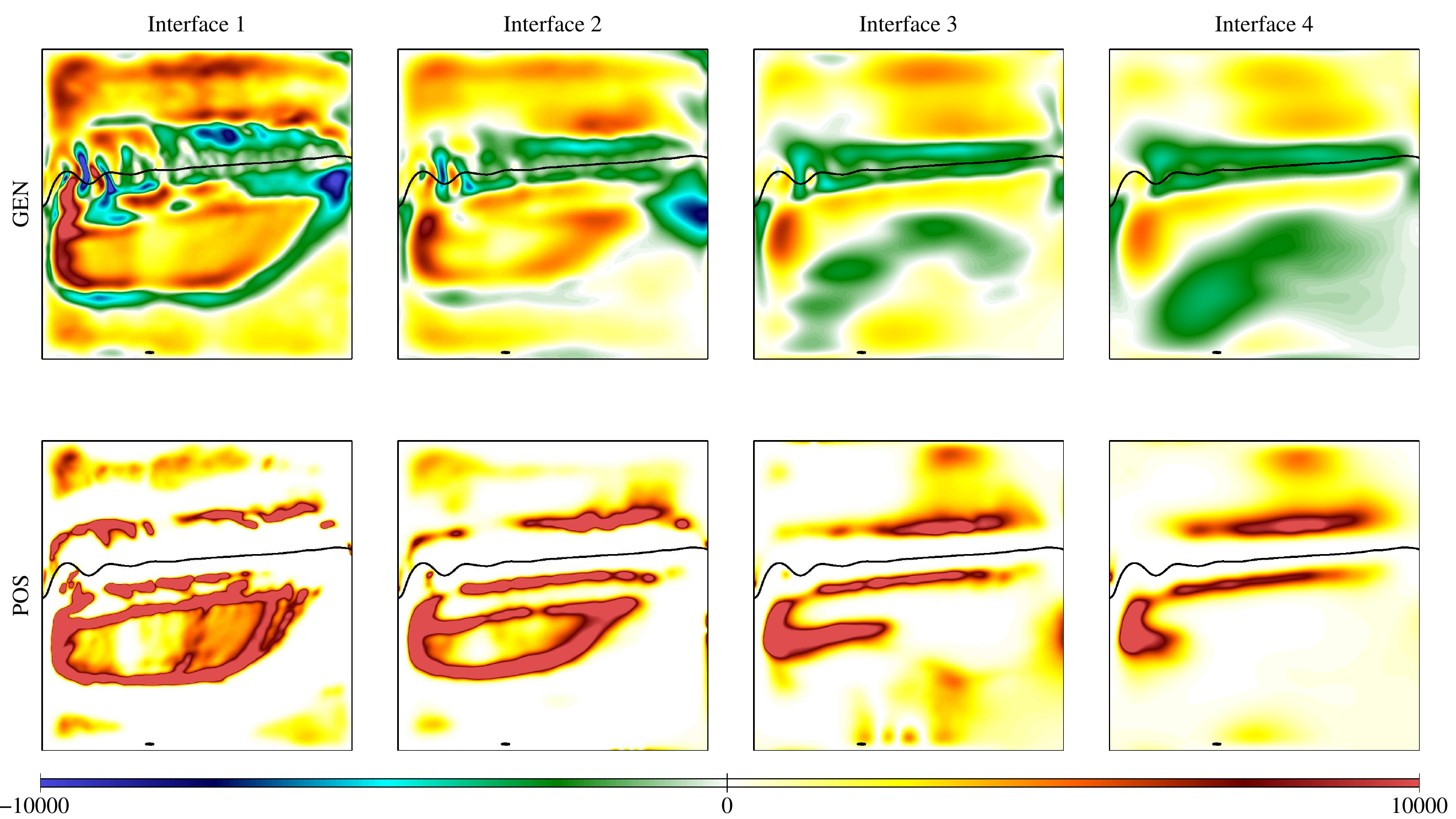}
	\caption{The diffusivity $\kappaGM$ (with units of $\mathrm{m^2}\
	\mathrm{s}^{-1}$) associated with the GEN case $\kappaGM = \xi$ (top layer)
	and POS case $\kappaGM = \xi^2$ (bottom layer) for buoyancy mixing over the
	four interfaces (column). The colour scale is saturated to show the spatial
	structures. Note that the GEN solution has not converged at the target global
	roughness of 7500 (see equation \ref{eq:GMroughness}), but instead the last
	converged solution at $\kappa^r_{\scriptsize\mbox{gm}} \approx 1100$ is
	displayed here.}
	\label{fig:kappa_archer_gm_L2}
\end{center}
\end{figure}

\begin{figure}[tbhp]
\begin{center}
  \includegraphics[width=0.9\textwidth]{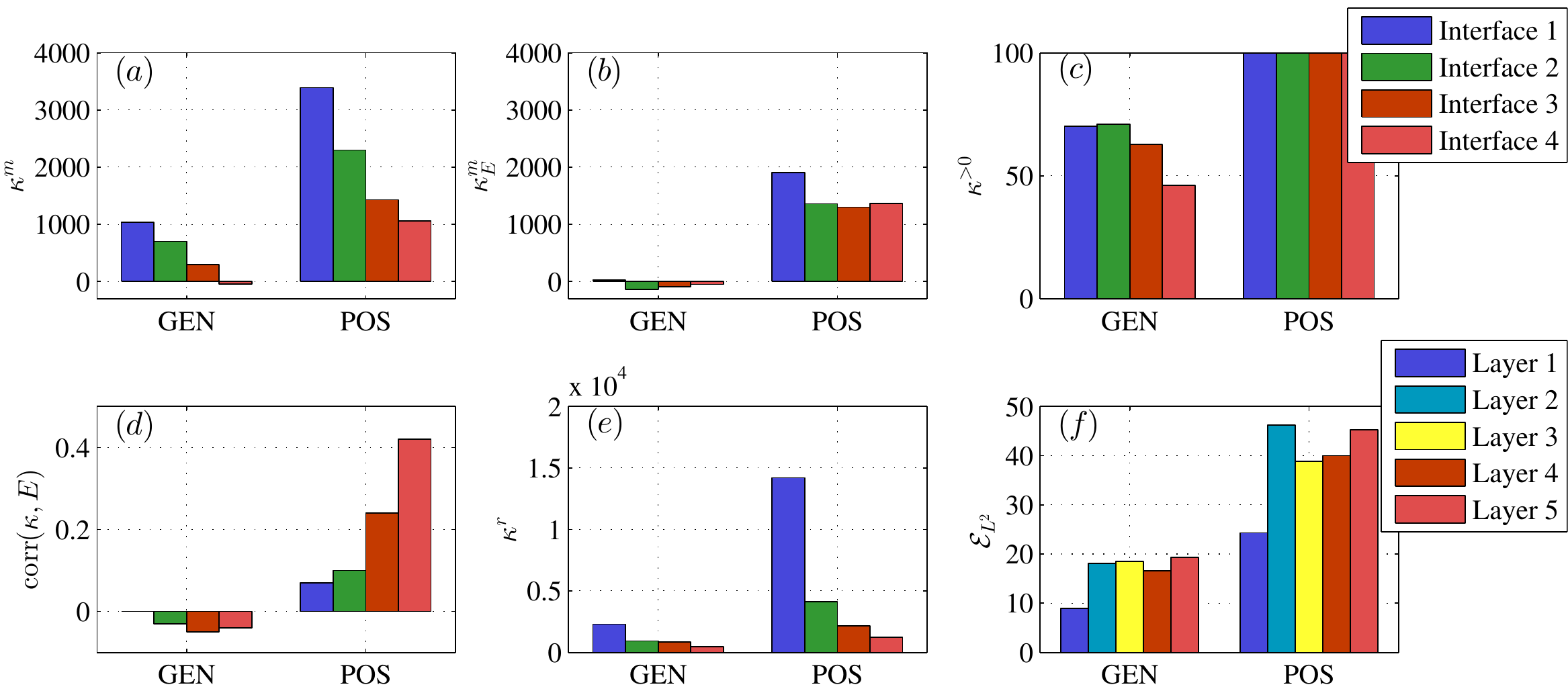}
  \caption{Bar graph comparing the diagnostic data across the four interfaces
  and five layers for the GEN and POS case associated with buoyancy mixing
  parameterisations: ($a,b$) the mean $\kappa^m$ and the eddy energy weighted
  mean $\kappa^m_E$ from equation \eqref{eq:kappa_mean}; ($c,d$) the positivity
  index $\kappa^{>0}$ and eddy energy weighted positivity index $\kappa^{>0}_E$
  from equation \eqref{eq:kappa_pos}; ($e$) the roughness $\kappa^r$ from
  equation \eqref{eq:kappa_rough}; ($f$) the relative $L^2$ error
  $\mathcal{E}_{L^2}$ from equation \eqref{eq:L2_err}. Note here is that the GEN
  case did not converge around the target global roughness of
  $\kappa^r_{\scriptsize\mbox{gm}} = 7500$ (see equation \ref{eq:GMroughness});
  instead, the last converged solution at $\kappa^r_{\scriptsize\mbox{gm}}
  \approx 1100$ is displayed here.}
  \label{fig:kappa_archer_gm_L2_bar}
\end{center}
\end{figure}

In the five-layer configuration there are two interfaces which are free from the
direct influence of upper layer forcing and bottom drag. These internal
interfaces are therefore likely to show a signal more consistent with the
quasi-adiabatic ocean interior, and hence more likely to correlate with the
action of down-gradient buoyancy mixing. The resulting $\kappaGM$ for the GEN
and POS case are shown in shown in Figure~\ref{fig:kappa_archer_gm_L2}. The
relevant diagnostic quantities are summarised in
Figure~\ref{fig:kappa_archer_gm_L2_bar}.

The first thing to note here is that the GEN case did not converge around the
target global roughness of $\kappa^r_{\scriptsize\mbox{gm}} = 7500$; instead,
the last converged solution at $\kappa^r_{\scriptsize\mbox{gm}} \approx 1100$ is
displayed here. This fact is perhaps noticeable in that the diagnosed
diffusivity is rather smooth, certainly compared to the three-layer case
displayed in Figure~\ref{fig:kappa_gm_vary_variety_L2}($a$). However, the
physical features are still robust, with the presence of the negative signal
down-stream of the mean jet across all interfaces, and the positive diffusivity
in the southern gyre, the western boundary and in the north-west
region. The mean is now largely positive except in the lowest interface, where
it is mildly negative. The eddy energy weighted mean however is mostly negative
and small in magnitude. The positivity index shows the diffusivity generally
takes positive values over the domain, although clear negative signals around
the mean jet are observed. The correlation between the diffusivity and the eddy
energy is generally small and negative. The roughness varies over the interfaces
because of the layer weighting employed in the definition of the global
roughness $\kappa^r_{\scriptsize\mbox{gm}}$. Given the resulting calculation
possesses a rather low roughness, the relative $L^2$ mis-match is respectable,
at less than $20\%$ over all fiver layers. The upper layer again has the lowest
mis-match.

For the POS case a non-zero solution is now found over all interfaces, in marked
contrast the earlier three layer case. The locations of positive signals again
correlate well with the locations of the positive signals observed in the GEN
case. The locations of low diffusivity in the POS case also correlate well with
the locations of negative signals in the GEN case. The unweighed mean and
especially the eddy energy weighted mean are by construction positive --- the
latter has a value of order $1300\ \textrm{m}^2\ \textrm{s}^{-1}$. The
correlation between the diffusivity and the eddy energy is larger than in the
GEN case. However, given that the POS case has a higher parameter roughness
(noting the less rough GEN case solution), the associated mis-match is still
significantly larger. This suggests that negative $\kappaGM$ is required for an
accurate match between the target and the parameterised eddy force functions.


\section{Conclusions}

A new method for diagnosing eddy diffusivities in a gauge-invariant fashion,
independent of dynamically inactive rotational flux components, has been
presented. This is achieved by seeking to match diagnosed and parameterised eddy
force functions through a one-shot optimisation procedure. The eddy force
function depends only upon eddy flux divergences, through an inverse elliptic
operator, and hence the force function is inherently smooth and non-local. The
optimisation problem allows control over the roughness of the resulting
diffusivity field. Combined, this method yields an optimal diffusivity that is
gauge-invariant, non-local and has controlled smoothness.

The approach has been applied to multi-layer quasi-geostrophic ocean gyre
simulations. Results have been shown here for data obtained from a three-layer
finite element simulation and a five-layer higher resolution finite difference
simulation. The diagnostic method has been applied for down-gradient PV mixing
and buoyancy mixing with a general unconstrained diffusivity and a positive
semi-definite diffusivity. The resulting optimality systems were implemented
using the FEniCS automated code generation system. Here the code generator
greatly facilitates parameterisation testing, as new methods can be implemented
and tested via small code modifications, and these changes are propagated
automatically. In particular, cost function Jacobians and Hessians are
formulated automatically via high-level algorithmic differentiation, and
specific code for the assembly of these discrete operators is generated
automatically.

Regarding down-gradient PV mixing parameterisations, the key conclusions are
that: (i) there are robust locally negative diffusivities that are present even
in the absence of rotational fluxes, although the mean diffusivity over the
horizontal domain is positive; (ii) the optimisation has success in matching the
eddy force function diagnosed from an eddying calculation in the lower layers,
but has less success in the upper layer where there is strong wind forcing
present; (iii) the locations of closed mean recirculations often correlate with
signals of positive diffusivity; (iv) there is positive correlation between the
eddy energy and the diffusivity in the lower layers. 

For down-gradient buoyancy mixing, negative signals are again present, although
in this instance some of this is attributed to the lack of vertical resolution
in the three-layer calculation. The five-layer calculations indicate
predominantly positive eddy diffusivities away from the location of the mean
jet, albeit with some strong negative signals in the southern part of the lower
two interfaces. Notwithstanding this exception, this is consistent with the
action of down-gradient momentum transfer input by the wind through the action
of baroclinic instability. However, within the mean jet, and particularly in the
lower layer and down-stream jet regions, the eddy diffusivity is strongly
negative, suggesting local baroclinic stability, and forcing of the mean jet
baroclinicity by the eddy buoyancy fluxes. These results are consistent with the
earlier observations reported in \cite{Berloff05a} and \cite{Maddison-et-al15}.

Throughout this paper the mis-match measure was defined via an $L^2$ norm,
measuring the mis-match between the diagnosed and parameterised eddy force
functions. Additional calculations were performed using a $H^1_0$ mis-match
measure, which measures the mis-match between diagnosed and parameterised
divergent eddy fluxes. The solutions obtained from the $H_0^1$ based mis-match
norms result in higher relative $L^2$ mis-matches; this may be attributed to the
fact that the $H^1_0$ case places more emphasis on the local, small-scale
features over the global, large-scale features. Calculations which attempted to
directly match diagnosed and parameterised eddy flux divergences (respectively,
$\grad^2 \Psi_e$ and $\grad\cdot(-\kappa\grad\overline{q})$) were not
successful.

It is possible to consider diagnostics which seek diffusivities which are
themselves defined in terms of the eddy energy \citep[e.g.,][]{Rodi87,
EdenGreatbatch08}. For example, one could consider the definition $\kappa =
\sqrt{E} \xi$, where $\xi$ a mixing length, or alternatively $\kappa = E \xi$,
where here $\xi$ is a time-scale. Via either of these approaches a given
roughness in the underlying parameter $\xi$ permits an increased roughness in
$\kappa$; that is, the eddy energy may be used to provide additional information
on the spatial structure of the diffusivity. Such diagnostics have been
investigated (not shown) and yield a root-mean-square mixing length of
$15$--$40~\mathrm{km}$, and a root-mean-square time scale of $3$--$10$~days. The
latter time-scale is similar to that described in \cite{McWilliamsGent94} for an
eddy kinetic energy dependent variant of GM with a spatially varying
coefficient.

For the purposes of eddy parameterisation, the diagnosed PV diffusivities
exhibit some desirable features. The mean diffusivity (either unweighted or eddy
energy weighted) is positive, and is generally also locally positive
(notwithstanding some regions of strong negative diffusivity, particularly in
the upper layer). The positive correlation of the diffusivity with eddy energy,
while somewhat modest, provides some additional support to the principle of eddy
energy based eddy parameterisations, for example as discussed in,
\cite{EdenGreatbatch08}, \cite{Cessi08}, \cite{MarshallAdcroft10} and
\cite{JansenHeld14}. Enforcing a positive semi-definite diffusivity leads to an
increased error at the selected parameter roughness. Indeed this latter approach
generally leads to a similar spatial diffusivity pattern as obtained with an
unconstrained diffusivity, but with negative diffusivities deleted, and somewhat
larger positive diffusivities elsewhere. 

Diagnosed interfacial buoyancy diffusivities are, for the purposes of eddy
parameterisation, potentially more problematic. In particular, at least in the
five layer calculation, where the influence of forcing and dissipation is weaker
for the intermediate layers, there are large scale and large magnitude negative
diffusivity signals, particularly in the region of the mean jet. While there are
also strong postive signals away the jet, the eddy energy weighted mean and the
correlation with the eddy energy are both negative. Enforcing a positive
semi-definite diffusivity again generally leads to a similar pattern of positive
spatial diffusivies, with larger magnitude, and with negative signals deleted.
The mis-match in this latter case, at the selected roughness, is also
significantly increased.

There do appear to be some robust structures appearing in the diagnosed
potential vorticity diffusivities. The five layer calculations suggest signals
broadly consistent with down-gradient diffusivities, hinting that such a closure
may be tractable here. This observation comes with the caveat that, in general,
a down-gradient potential vorticity closure violates momentum conservation
\citep[e.g.,][]{Marshall81, Marshall-et-al12} through the failure to preserve
the underlying tensorial structure resulting from the Taylor-Bretherton identity
\citep[e.g.,][]{Griffies-ocean, PopovychBihlo12, MaddisonMarshall13}.

The diagnosed buoyancy diffusivities imply a region near the mean jet with a
robust negative diffusivity imply a region with robust negative diffusivity,
consistent with the action of baroclinic stability. This suggests that, at least
in the regions of strong lateral shear, a closure for eddy buoyancy fluxes
should permit a degree of backscatter. This is not typically captured within
current eddy parameterisation schemes.

The optimisation procedure may potentially be extended to the primitive
equations, provided an analogous eddy force function may be defined. One could
consider, for example, a force function defined as in \citet{MarshallPillar11}.
A practical limitation here is likely to be the difficulty of solving the
associated ill-conditioned optimality systems. In this article this was
addressed by reducing the size of the problems, through interpolation onto a
coarser mesh, combined with the use of direct solvers which are practical for
these problem sizes. For larger problems more advanced methods, such as the use
of iterative methods with appropriate pre-conditioners for the relevant linear
systems, are likely to be required.

In practice a parameterisation quality may not necessarily be determined by its
ability to represent eddy statistics themselves. That is, it may be acceptable
for a given parameterisation to imply a differing eddy diffusivity if it
nevertheless yields a high quality mean state. A more advanced method of
diagnosing eddy diffusivities, for example, could seek to invert for a
diffusivity which yields an optimal mis-match between high resolution and
parameterised mean states. Such a diagnostic would apply the dynamical equations
themselves as a constraint on the optimisation, and replace the unconstrained
cost function used here with a mis-match measure defined in terms of the
deviation of the parameterised model from the target high resolution reference.


\section*{Acknowledgements}

This work was funded by the UK Natural Environment Research Council grant
NE/L005166/1. This work used the ARCHER UK National Supercomputing Service
(\verb!http://www.archer.ac.uk!). The parallel finite difference code used in this
article is based upon an earlier code provided by P. S. Berloff \citep[see
also][]{Karabasov-et-al09}. The data used for generating the plots in this
article is available through the Edinburgh DataShare service at
\verb!http://dx.doi.org/10.7488/ds/366!.





\bibliographystyle{elsarticle-harv} 
\bibliography{refs}


\end{document}